\documentstyle[aaspp4]{article}

\begin{document}

\title{The Spectral Signature of Dust Scattering and Polarization in the Near 
IR to Far UV. I. Optical Depth and Geometry Effects}

\author{Victor G. Zubko\altaffilmark{1} and Ari Laor}
\affil{Department of Physics, Technion -- Israel Institute
             of Technology, Haifa 32000, Israel}

\altaffiltext{1}{On leave from the Main Astronomical Observatory,
                 National Academy of Sciences of the Ukraine, Kiev}

\begin{abstract}
Spectropolarimetry from the near IR to the far UV of light
scattered by dust provides a valuable diagnostic of the dust composition,
grain size distribution and spatial distribution. To facilitate the use
of this diagnostic,
we present detailed calculations of the intensity and polarization
spectral signature of light scattered by optically thin and 
optically thick dust in various geometries. 
The polarized light radiative transfer calculations are carried out
using the adding-doubling method for a plane-parallel slab, and 
are extended to an optically thick sphere by integrating over its surface.
The calculations are for the Mathis, Rumple \& Nordsieck
Galactic dust model, and cover the range from 1~$\mu m$ to 500~\AA.

We find that the wavelength dependence of the scattered light intensity 
provides a sensitive 
probe of the optical depth of the scattering medium, while the polarization 
wavelength dependence provides a probe of the grain scattering properties, 
which is practically independent of optical depth. We provide a detailed set of 
predictions, including polarization maps, which can be used to probe 
the properties of dust through imaging spectropolarimetry in the near IR
to far UV of various Galactic and extragalactic objects. 
In a following paper we use the codes developed here to 
provide predictions for the dependence of the intensity and polarization 
on grain size distribution and composition.

\end{abstract}


\section{Introduction}
       \label{sec:intr}

Light scattering induces polarization, and the intensity and polarization of
the scattered light tells us the properties of the scattering medium. 
Continuum radiation can be scattered over large volumes of space
either by free electrons, or by dust grains in either diffuse or compact
clouds.
Electron scattering is wavelength independent (below X-ray energies) 
and follows a simple scattering phase function 
($\propto 1+\cos^2\Theta_{\rm scat}$), but dust scattering is much more
complicated since the scattering efficiency and the phase function can be
strongly dependent on wavelength, grain size, and grain composition.
Thus, light scattered by dust carries valuable information on the dust
properties, and the purpose of this paper is to facilitate the extraction
of this information.

Dust reflected light is observed in various Galactic objects, either from large 
and diffuse reflection nebulae
(e.g. Witt et al. \cite{w92}, \cite{w93}; Gordon et al. \cite{g94};
Calzetti et al. \cite{c95}), or from more compact, but still
spatially resolved objects, such as dense molecular clumps (e.g. the Eagle 
nebula, Hester et al. 1996),
protoplanetary disks (the Orion nebula, O'Dell, Wen, \& Hu
\cite{o93}; Bally et al. \cite{b98}), cometary knots in planetary
nebulae (Balick et al. \cite{ba98}; Burkert \& O'Dell \cite{bu98}),
and bipolar nebulae (the Egg nebula, Sahai et al. 1998a,b).
Dust reflection is also inferred in some compact unresolved objects 
(e.g. Schulte-Ladbeck et al. 1992; Rees \& Sitko 1996), and is also
likely to be present in young stellar objects with circumstellar disks
(e.g. Chiang \& Goldreich 1997).  

Dust reflected light is also observed in various extragalactic objects.
In particular in active galaxies, where dust obscuration plays a key
role in unification schemes (Antonucci 1993; Urry \& Padovani 1995).
Specific evidence for scattering is present in Seyfert 2 galaxies
(e.g. Antonucci \& Miller \cite{a85}; Miller, Goodrich \& Mathews \cite{m95};
Tran 1995a,b; Capetti et al. \cite{c96}), radio galaxies (e.g. Jannuzi et al.
\cite{j95}; Dey et al. \cite{d96}; Cimatti et al. \cite{c98}; 
Tran et al. \cite{t98}; Hurt et al. \cite{h99}), broad absorption line quasars
(e.g. Cohen et al. \cite{c95}; Goodrich \& Miller \cite{g95};
Schmidt \& Hines et al. \cite{s99}), red quasars (e.g. Brotherton et al.
\cite{b98}; De Breuck et al. \cite{d98}), and ultraluminous IR active galaxies
(e.g. Wills et al. \cite{wi92}; Hines \& Wills \cite{hi93};
Hines et al. \cite{hi99}). In some of these objects the  
polarization rises towards shorter wavelengths, strongly suggesting optically
thin dust scattering. 

Despite its potential usefulness, almost no data currently exists on 
spectropolarimetry of spatially resolved scattered light in Galactic 
objects (see a rare exception in Code et al. 1996). 
Fortunately, significant amounts of spectropolarimetry,
though mostly spatially unresolved, are available
for extragalactic objects, in particular at the rest frame UV (for high
$z$ objects), where most of the spectral polarization features are predicted, 
and which therefore provides the strongest diagnostic power. 

In this paper we provide a detailed set of predictions for
the wavelength dependence of the intensity, percent polarization, polarization 
plane, and circular polarization, of light scattered by dust in the near IR to
the far UV. These properties are explored here for a given dust model over a 
range of scattering 
geometries for either optically thin or optically thick dust. In a following 
paper we explore these properties for various dust models, as characterized
by the grain size distribution and grain composition.
These predictions, together with new high quality spectropolarimetry in the near
IR to the far UV of reflection dominated objects, should allow significant
constraints on the properties of dust in various Galactic and extragalactic
environments.
  
Earlier calculations of polarized radiative transfer in a dusty optically
thick medium were made by White (1979a,b), Kartje (\cite{kar95}),
Code \& Whitney (\cite{code-whitney}), and Wolf \& Henning (\cite{wh99};
see also optically thin calculations for radio galaxies in Manzini \& di 
Serego Alighieri 1996).
The calculations presented here improve on these earlier calculations 
either by the use of an accurate radiative transfer, rather than a Monte Carlo
calculation, or by using updated dielectric functions of silicate and graphite
grains, by using the actual, rather than approximated scattering phase matrix,
by extending our results down to 500 {\AA} (relevant for high redshift extra 
galactic objects), and by the detailed set of results presented. 

The paper is organized as follows. In \S 2 we provide a detailed description 
of the theoretical approach. The results for scattering by optically thin dust,
and by an optically thick slab and sphere are presented in \S 3,
and the paper is summarized in \S 4. A detailed description of
the numerical implementation is provided in the Appendix, where we also
provide tests of the codes based on a comparison with accurate analytic
results for electron scattering, and with earlier numerical results for
dust scattering. We also show in the Appendix a comparison with the much 
simpler case of an electron scattering sphere.

\section{Theoretical approach}
       \label{sec:theor_appr}

The main purpose of this paper is to study in detail the wavelength
dependence of scattering and
polarization of light incident on optically thick dust in simple
geometrical configurations. 
The distance from the illumination source to the dusty nebula is assumed
here to be much larger than the characteristic length corresponding to an 
optical depth $\tau\sim 1$ within the dusty nebula, so one may use the parallel 
illumination assumption. 
Dusty slabs or spheres are the simplest possible dust configurations
to consider, and are relevant for the various optically thick scattering
objects mentioned in \S 1.

Radiative transfer in a plane-parallel slab is a standard problem in
the theory of radiative transfer (Chandrasekhar \cite{chandrasekhar60}).
To solve this problem we use
the adding-doubling method (Hovenier \cite{hov71}; Hansen \& Travis
\cite{ht74}; Evans \& Stephens \cite{ev91}), which is efficient
and numerically stable up to large optical depths.
We follow the formulation of the adding-doubling method
proposed by Evans \& Stephens (\cite{ev91}). 
To solve the case of an optically thick dusty sphere, 
we assume that the geometrical depth at which $\tau\sim 1$ is much smaller
than the radius of the sphere, so the  
surface of the sphere can be treated as a collection of a large number
of small plane-parallel slabs with a horizontal size much less
than the sphere radius, but still much larger than the vertical depth at
which $\tau\sim 1$. The scattering properties of the sphere are
obtained by
integrating the local solution for the slabs over the sphere surface.
Note that the surface integration approach can be applied to practically 
any optically thick geometry.

Light scattering by a sphere can also be calculated using the Monte Carlo 
method (Witt \cite{witt77}), as was done by Code \& Whitney
(\cite{code-whitney}). However, the efficiency of the Monte Carlo method 
drops steeply at large optical depths,
typically for $\tau >$ 5--10 (Fischer, Henning, \& Yorke \cite{fhy94};
Code \& Whitney \cite{code-whitney}; Wolf \& Henning \cite{wh99}), and it
does not allow the accuracy provided by an exact numerical radiative transfer
for a large optical depth.
However, the disadvantage of the numerical scheme employed here for a sphere
is that it applies only for $\tau \gg 1$, unlike the  
Monte Carlo method which allows one to
treat spheres with intermediate optical depth.

In the following subsections we provide a detailed description 
of our theoretical approach. We first briefly review the basics
of single scattering by ensembles of spherical grains (section
\ref{subsec:single_scat}), introducing the phase
matrix, albedo, and other parameters relevant for multiple
scattering. In section \ref{subsec:mult_scat} we describe the polarized radiative 
transfer in a slab, which provides the basis for this and forthcoming studies.
In section \ref{subsubsec:math_formul} we formulate the problem
and introduce the main parameters. We next derive the expression for
the scattering matrix (section \ref{subsubsec:scat_mat}) in a form 
suitable for efficient implementation of the adding-doubling method,
which is described in section \ref{subsubsec:doubl_add}.
In section \ref{subsec:scat_sphere}, we describe how to calculate
the scattering characteristics of an optically thick sphere using
the solution for a slab.

  \subsection{Single scattering by dust grains}
       \label{subsec:single_scat}

In order to solve the radiative transfer in an extended dusty medium,
it is necessary to obtain first the scattering properties per 
unit volume. If the mean distance between
dust particles in the medium considerably exceeds the typical grain size,
and the particles are randomly distributed (as applies for cosmic dust),
then there are no coherence effects and the particles scatter the light 
independently. The scattering also does not 
change the frequency. The transfer in larger, optically thick
volumes, is studied by solving the equations of radiative
transfer, using the results for the single scattering problem.
In this section we introduce the main parameters which describe single
scattering; the intensity vector,
the phase matrix and the single-scattering albedo. We next present specific
expressions of these parameters for the cases of Mie and Rayleigh scattering
relevant for the present study.
 
The intensity and polarization of a beam of light are fully
described by the 4-vector $\bf{I}$ of Stokes parameters $(I, Q, U, V)$
(Chandrasekhar \cite{chandrasekhar60}; Bohren \& Huffman \cite{bh83}),
defined with respect to a certain plane of reference.
The parameter $I$ describes the total light intensity, 
$Q$ and $U$ describe the linear polarization, frequently expressed
through the degree of linear polarization $p_{\rm{l}}=\sqrt{Q^2+U^2}/I$ and
the position angle of the direction of linear polarization with respect to
the local meridian plane $\theta_{\rm{p}}=\frac{1}{2} \arctan(U/Q)$, such that
$\cos 2\theta_{\rm{p}}$ has the same sign as $Q$. The last parameter,
$V$, describes the circular polarization, and the degree of circular polarization
is $p_{\rm{c}}=V/I$. Note that the intensity vector $\bf{I}$ is
actually {\em a pseudo-vector}, since it does not transform like a true vector
under a coordinate transformation. It is possible to build
the so-called radiation density matrix $\hat \rho$ in terms of the Stokes
parameters (Dolginov, Gnedin, \& Silant'ev \cite{dgs95})
\begin{equation}
\hat \rho = \frac{1}{2} \left[ \begin{array}{cc}
            I + Q   & U + iV\\
            U - iV  & I - Q
            \end{array} \right],
  \label{eq:hatrho}
\end{equation}
which is {\em a tensor} in coordinate space. However, we base
our calculations on the the intensity vector $\bf{I}$,
as is common for cosmic dust studies.

The scattering from a small volume is obtained by summing over the scatterings
of the grains within that volume.
We assume a model of spherical grains, which is frequently used in various 
applications. The approach we develop is applicable not only
for classical homogeneous spherical grains (Mie particles), but also
for spherical grains of multilayer or composite structure.

If a grain of radius $a$ is illuminated by parallel radiation
of intensity ${\bf I}_{\rm{in}}$ at a wavelength $\lambda$, 
then the intensity of the radiation scattered by the grain in the direction
$\Theta_{\rm{scat}}$ and measured at a distance
$r$ in the far-field ($r\gg\lambda$) is
\begin{equation}
{\bf I}_{\rm{out}}(\Theta_{\rm{scat}},a,\lambda) =
   {{\sigma_{\rm{sca}}}(a,\lambda) \over {4{\pi}}r^2} \;
   {\cal P}(\Theta_{\rm{scat}},a,\lambda) \; {\bf I}_{\rm{in}}
  \label{eq:I}
\end{equation}
where $\sigma_{\rm{sca}}(a,\lambda)$ is the scattering cross-section of
the grain and ${\cal P}(\Theta_{\rm{scat}},a,\lambda)$ is a 4-by-4 matrix
called the phase matrix.

Since energy can be removed from the incident radiation by scattering
as well as by absorption, it is important to introduce also
the absorption and extinction cross-sections: $\sigma_{\rm{abs}}$
and $\sigma_{\rm{ext}}=\sigma_{\rm{sca}}+\sigma_{\rm{abs}}$, respectively.
Then we may define the so-called single-scattering albedo
${\tilde \omega}=\sigma_{\rm{sca}}/\sigma_{\rm{ext}}$, which gives 
the probability that a photon hitting a grain will be scattered.
It is common to use the dimensionless efficiency
factors for scattering, absorption and extinction, defined
as ratios of respective cross-sections to the grain geometrical cross-section,
equal in our specific case to $\pi a^2$:
\begin{equation}
  Q_{\rm{sca}}={ \sigma_{\rm{sca}} \over {\pi a^2} }, \;\;\;\;\;\;
  Q_{\rm{abs}}={ \sigma_{\rm{abs}} \over {\pi a^2} }, \;\;\;\;\;\;
  Q_{\rm{ext}}={ \sigma_{\rm{ext}} \over {\pi a^2} },
  \label{eq:Qs}
\end{equation}
and thus
${\tilde \omega}=Q_{\rm{sca}}/Q_{\rm{ext}}$.

The phase matrix $\cal P$ contains the full information about
the angular distribution and state of polarization of the scattered
radiation. For spherical grains, $\cal P$ has an especially simple
expression (Van de Hulst \cite{vdh57}; Bohren \& Huffman \cite{bh83})
\begin{equation}
{\cal P}(\Theta_{\rm{scat}},a,\lambda) = \left[ \begin{array}{rrrr}
            P_1(\Theta_{\rm{scat}},a,\lambda) & P_2(\Theta_{\rm{scat}},a,\lambda) &   0  &0\\
            P_2(\Theta_{\rm{scat}},a,\lambda) & P_1(\Theta_{\rm{scat}},a,\lambda) &   0  &0\\
            0   &   0 & P_3(\Theta_{\rm{scat}},a,\lambda)  & P_4(\Theta_{\rm{scat}},a,\lambda)\\
            0   &   0 & -P_4(\Theta_{\rm{scat}},a,\lambda) & P_3(\Theta_{\rm{scat}},a,\lambda)
            \end{array} \right],
  \label{eq:P}
\end{equation}
where only three of the four matrix elements 
are independent since: $P_1^2=P_2^2+P_3^2+P_4^2$.
The scattering plane, defined by the incoming and outgoing
beams of light, is the plane of reference for the vectors of the Stokes
parameters. The element $P_1$ is the phase function, which
describes the probability that the unpolarized incident light will be
scattered in any specific direction. The phase matrix ${\cal P}$ is therefore
normalized such that
\begin{equation}
\int_{4{\pi}} P_1 \; { {{\rm{d}}{\Omega}} \over {4{\pi}} } = 1.
  \label{eq:P-norm}
\end{equation}
The angular dependence of the phase function $P_1$ can be characterized
by the so-called asymmetry parameter $g$:
\begin{equation}
g = \langle{\cos \theta}\rangle =
  \int_{4{\pi}} \cos {\theta} \; P_1 \; { {{\rm{d}}{\Omega}} \over {4{\pi}} },
  \label{eq:g_def}
\end{equation}
where $g$ ranges from --1 (complete backscattering) to 1 (complete
forward scattering) with $g=0$ corresponding to isotropic scattering.

    \subsubsection{Mie scattering}
       \label{subsubsec:mie_scat}

Light scattering by a spherical homogeneous particle of an arbitrary radius
$a$ is given by the Mie series solution (the formulae
presented here are valid also for the case of a concentric multilayer
sphere). The elements of the phase matrix for a Mie-scattering particle are
(Bohren \& Huffman \cite{bh83}):
\begin{equation}
\begin{array}{ll}
P_1 = {2 \over {x^2 Q_{\rm{sca}}}} (\mid S_2 \mid^2 + \mid S_1\mid^2), &
P_2 = {2 \over {x^2 Q_{\rm{sca}}}} (\mid S_2 \mid^2 - \mid S_1\mid^2),\\
P_3 = {2 \over {x^2 Q_{\rm{sca}}}} (S_2^{\ast} S_1 + S_2 S_1^{\ast}), &
P_4 = {{2i} \over {x^2 Q_{\rm{sca}}}} (S_2^{\ast} S_1 - S_2 S_1^{\ast}),
\end{array}
  \label{eq:P_Mie}
\end{equation}
where $x=2{\pi}a/\lambda$ is the size parameter, and $S_1$ and $S_2$ are
the scattering amplitudes, in general complex functions, for which
the Mie theory gives (Bohren \& Huffman \cite{bh83}):
\begin{equation}
S_1(\mu) = \sum_{k=1}^{\infty}{ {{2k+1} \over {k(k+1)} } [a_k {\pi}_k(\mu) + b_k {\tau}_k(\mu)]  },\;\;\; 
S_2(\mu) = \sum_{k=1}^{\infty}{ {{2k+1} \over {k(k+1)} } [a_k {\tau}_k(\mu) + b_k {\pi}_k(\mu)]  }.
  \label{eq:s1s2}
\end{equation}
where $\mu = \cos \Theta_{\rm{scat}}$, the functions $\pi_k(\mu)$ and
$\tau_k(\mu)$ contain the information about angular dependence,
and $a_k$ and $b_k$ are the so-called scattering coefficients.
The angular functions $\pi_k(\mu)$ and $\tau_k(\mu)$ can be calculated
from the simple recursion relations (Bohren \& Huffman \cite{bh83}):
\begin{equation}
\pi_{k+2} = {{2k+3} \over {k+1}} \mu \pi_{k+1} - {{k+2} \over {k+1}} \pi_k, \;\;\;
\tau_{k+2} = (k+2) \mu \pi_{k+2} - (k+3) \pi_{k+1},
  \label{eq:pi_tau}
\end{equation}
starting from $\pi_0$=0 and $\pi_1$=1. The coefficients $a_k$ and $b_k$,
which are generally complex, characterize the contribution of various
multipole modes to the scattering process.
These are dependent on $a$, $\lambda$ and the grain material complex
refractive index $n=n_r + i n_i$. The specific expressions for
$a_k$ and $b_k$, which are also calculated with recursion relations,
may be found, e.g. in Bohren \& Huffman (\cite{bh83}).

Once $a_k$ and $b_k$ are known, the efficiency factors $Q_{\rm{sca}}$
and $Q_{\rm{ext}}$, and the asymmetry parameter $g$, are given by
(Bohren \& Huffman \cite{bh83}):
\begin{equation}
Q_{\rm{sca}} = {2 \over x^2} \sum_{k=1}^{\infty} (2k+1) (\mid a_k \mid^2 + \mid b_k \mid^2), \;\;\;
Q_{\rm{ext}} = {2 \over x^2} \sum_{k=1}^{\infty} (2k+1) \;\Re \{a_k + b_k\},
  \label{eq:Q_mie}
\end{equation}
\begin{equation}
g = {4 \over {x^2 Q_{\rm{sca}}}}
    \sum_{k=1}^{\infty} \left[ { {k(k+2)} \over {k+1} } \;\Re\{a_k a_{k+1}^{\ast} + b_k b_{k+1}^{\ast}\} +
            { {2k+1} \over {k(k+1)} } \;\Re\{a_k b_k^{\ast}\} \right].
  \label{eq:g_mie}
\end{equation}
Computational implementation
of formulae (\ref{eq:P_Mie}--\ref{eq:g_mie}) can be found in
Bohren \& Huffman (\cite{bh83}) and Wiscombe (\cite{wis79}).

    \subsubsection{Rayleigh scattering}
       \label{subsubsec:rayleigh_scat}

The Rayleigh scattering approximation applies when the grains are very small, i.e.
$x \ll 1$, and in addition $x$$\mid$$n$$\mid\; \ll 1$. The above conditions mean
that the particles interact with the radiation mainly as electric dipoles
with a negligibly small contribution of higher multipoles.
The elements of the phase matrix in this case can be obtained
from those for Mie scattering [eqs (\ref{eq:P_Mie})] by retaining in
the expansions (\ref{eq:s1s2}) the terms responsible for the electric
dipole interaction, which gives the very simplified expressions 
(Bohren \& Huffman \cite{bh83}):
\begin{equation}
\begin{array}{ll}
P_1 = {3 \over 4} (1 + \cos^2 \Theta_{\rm{scat}}), &
P_2 = - {3 \over 4} \sin^2 \Theta_{\rm{scat}},\\
P_3 = {3 \over 2} \cos \Theta_{\rm{scat}}, &
P_4 = 0.
\end{array}
  \label{eq:P_Ray}
\end{equation}
The phase matrix $\cal P$ is a function of 
the scattering angle $\Theta_{\rm{scat}}$ only, unlike the phase matrix
for Mie scattering particles which depends also on grain properties:
$a$, $n$, and $\lambda$.
The expressions for the optical efficiencies also become simpler:
\begin{equation}
Q_{\rm{sca}} = {8 \over 3} x^4 \left|{{n^2-1} \over {n^2+2}}\right|^2, \;\;\;
Q_{\rm{ext}} = 4x \; Im\left\{{ {n^2-1} \over {n^2+2} }\right\},
  \label{eq:Q_ray}
\end{equation}
where $Im$ stands for the imaginary part.
Note that for Rayleigh scattering by free electrons
(Thomson scattering) the elements of the phase matrix are the same
as in (\ref{eq:P_Ray}).

    \subsubsection{Single scattering by a mixture of size-distributed grains}
       \label{subsubsec:mixt_scat}

In the previous sections we considered light scattering by a single
grain. A dust mixture contains grains made up of various species 
and a range of sizes. Let $f^i(a)$ be the size distribution function 
for the $i$th dust species, that is the number density of grains of type 
$i$ having sizes
within the interval $a$ to $a$+d$a$.
The scattering parameters of a volume element is obtained
by averaging the single grain scattering over the dust
species and the grain-size distributions, which yields the following expressions:
\begin{equation}
P_j(\Theta_{\rm{scat}}, \lambda) = {
   \frac
   {
     \sum_i \int_0^{\infty} {\pi}a^2 \; Q_{\rm{sca}}^i(a,\lambda) \;
        P_j^i(\Theta_{\rm{scat}},a,\lambda) \; f^i(a) \; {\rm{d}} a
   }
   {
     \sum_i \int_0^{\infty} {\pi}a^2 \; Q_{\rm{sca}}^i(a,\lambda) \;
        f^i(a) \; {\rm{d}} a
   }
  },
  \label{eq:P_ave}
\end{equation}
\begin{equation}
Q_{\rm{sca}}(\lambda) = {
   \frac
   {
     \sum_i \int_0^{\infty} {\pi}a^2 \; Q_{\rm{sca}}^i(a,\lambda) \;
        f^i(a) \; {\rm{d}} a
   }
   {
     \sum_i \int_0^{\infty} {\pi}a^2 \; f^i(a) \; {\rm{d}} a
   }
  },
  \label{eq:Q_sca_ave}
\end{equation}
\begin{equation}
Q_{\rm{ext}}(\lambda) = {
   \frac
   {
     \sum_i \int_0^{\infty} {\pi}a^2 \; Q_{\rm{ext}}^i(a,\lambda) \;
        f^i(a) \; {\rm{d}} a
   }
   {
     \sum_i \int_0^{\infty} {\pi}a^2 \; f^i(a) \; {\rm{d}} a
   }
  },
  \label{eq:Q_ext_ave}
\end{equation}
\begin{equation}
g(\lambda) = {
   \frac
   {
     \sum_i \int_0^{\infty} {\pi}a^2 \; Q_{\rm{sca}}^i(a,\lambda) \;
        g(a,\lambda) \; f^i(a) \; {\rm{d}} a
   }
   {
     \sum_i \int_0^{\infty} {\pi}a^2 \; Q_{\rm{sca}}^i(a,\lambda) \;
     f^i(a) \; {\rm{d}} a
   }
  },
  \label{eq:g_ave}
\end{equation}
\begin{equation}
{\tilde \omega} = {
   \frac
   {
     \sum_i \int_0^{\infty} {\pi}a^2 \; Q_{\rm{sca}}^i(a,\lambda) \;
        f^i(a) \; {\rm{d}} a
   }
   {
     \sum_i \int_0^{\infty} {\pi}a^2 \; Q_{\rm{ext}}^i(a,\lambda) \;
        f^i(a) \; {\rm{d}} a
   }
  },
  \label{eq:alb_ave}
\end{equation}
where the sum index $i$ runs over all grain species.
It is useful to define the scattering and extinction coefficients:
\begin{equation}
k_{\rm{sca}}(\lambda) = \sum_i \int_0^{\infty} {\pi}a^2 \; Q_{\rm{sca}}^i(a,\lambda) \;
     f^i(a) \; {\rm{d}} a ,
  \label{eq:k_sca}
\end{equation}
\begin{equation}
k_{\rm{ext}}(\lambda) = \sum_i \int_0^{\infty} {\pi}a^2 \; Q_{\rm{ext}}^i(a,\lambda) \;
     f^i(a) \; {\rm{d}} a ,
  \label{eq:k_ext}
\end{equation}
measured in units of length$^{-1}$. The inverse quantities $k_{\rm{sca}}^{-1}$
and $k_{\rm{ext}}^{-1}$ represent the mean free path 
of photons for scattering and extinction.

  \subsection{Scattering by an optically thick dusty slab}

       \label{subsec:mult_scat}

In this section we formulate the solution for polarized radiative transfer in a 
a plane-parallel dusty slab. Since dust
scattering does not change $\lambda$, 
we omit the explicit dependence of the intensity and polarization on 
$\lambda$.

    \subsubsection{Mathematical formulation of the problem}
       \label{subsubsec:math_formul}

Let a dusty slab of optical thickness $\tau_0$ be illuminated
by unpolarized radiation incident from
a direction $\theta_0$ with a flux $F_0$ (see Fig.
\ref{fig:coord_slab}). The equation of polarized radiative
transfer for the slab is (Chandrasekhar \cite{chandrasekhar60}):
\begin{equation}
\mu {{{\rm{d}}{\bf I}(\tau, \mu, \Phi)} \over {{\rm{d}}\tau}} =
    -{\bf I}(\tau, \mu, \Phi)
    + {\bf F}(\tau, \mu, \Phi, {\bf I}) + {\bf S}(\tau, \mu, \Phi),
      \label{eq:rad_tran}
\end{equation}
\begin{equation}
{\bf F}(\tau, \mu, \Phi, {\bf I}) =
    {{\tilde \omega} \over {4\pi}}
    \int_0^{2\pi} \!\!  \int_{-1}^1 \!\! {\cal F}(\mu,\Phi,{\mu}',{\Phi}') \;
    {\bf I}(\tau, {\mu}', {\Phi}') \; {\rm{d}}{\mu}' \; {\rm{d}}{\Phi}',
      \label{eq:F_int}
\end{equation}
\begin{equation}
{\bf S}(\tau, \mu, \Phi) =
    {{\tilde \omega} \over {4\pi}} \; F_0 \; e^{-{{\tau} \over {\mu_0}}} \;
   {\cal F}(\mu,\Phi,{\mu}_0,{\Phi}_0) \; {\bf 1}
    + (1-{\tilde \omega}) \; B(T) \; {\bf 1},
      \label{eq:S}
\end{equation}
where ${\bf I}$ is the intensity vector of the diffuse radiation
at an optical depth $\tau$, going along the direction ($\theta$, $\Phi$)
or ($\mu$, $\Phi$); where $\mu$ is the cosine of zenith angle $\theta$,
measured from the upper normal for the upward intensity and from the downward
normal for the downward intensity (see Fig. \ref{fig:coord_slab});
$\Phi$ is the azimuthal angle, measured counterclockwise 
from the projection of the incoming central source beam
onto the slab. Note that all scattering quantities depend on the azimuth
difference rather than on the azimuth itself. For
the incoming beam $\Phi_0=0$ and $\mu_0=\cos \theta_0$.

${\bf F}$ is the scattering integral, expressing the angular
transformation of the diffuse radiation due to scattering.
$\cal F$ is the 4-by-4 scattering matrix that relates the intensities
of incoming and outgoing radiation, taken in their specific reference
planes (see section \ref{subsubsec:scat_mat} for more details).

${\bf S}$ is the source function 4-vector, which includes a `pseudo-source'
of single scattered radiation of the central source and
unpolarized thermal emission of dust. The latter may be
approximated by a black-body radiation $B$ of temperature $T$.
${\bf 1}$ is the unity 4-vector (1,~0,~0,~0). Note that in this study
we are interested in light scattering in the UV and the optical,
where the contribution of the dust thermal emission is 
negligibly small. We therefore exclude this effect below.

To complete the formulation of the problem, we should specify
the boundary conditions for the intensity on the top and bottom
faces of the slab (Fig. \ref{fig:coord_slab}):
\begin{equation}
{\bf I}(0, \mu, \Phi) = {\bf I}_{\rm{t}}^{\downarrow}(\mu, \Phi), \;\;\;
{\bf I}(\tau_0, \mu, \Phi) = {\bf I}_{\rm{b}}^{\uparrow}(\mu, \Phi),
      \label{eq:boundary}
\end{equation}
where ${\bf I}_{\rm{t}}^{\downarrow}$ (${\bf I}_{\rm{b}}^{\uparrow}$) is
the intensity of diffuse radiation from external sources 
incident on the top (bottom) of the slab.
For a slab with a large optical thickness $\tau_0 \gg 1$, instead of
the second boundary condition in (\ref{eq:boundary}) we expect that
the intensity ${\bf I}(\tau, \mu, \Phi)$ at $\tau \to \infty$
be limited. Note that in this study the intensities
${\bf I}_{\rm{t}}^{\downarrow}$ and ${\bf I}_{\rm{b}}^{\uparrow}$
of external diffuse radiation in the optical/UV are expected to be 
negligibly small compared to
the intensity of the central source, so we may set them to zero.
However, we retain ${\bf I}_{\rm{t}}^{\downarrow}$ and
${\bf I}_{\rm{b}}^{\uparrow}$ until the end of this section
for purpose of generality.

In the following sections we describe how the problem
of radiative transfer in a dusty slab can be solved efficiently 
by the doubling and adding method (Hovenier \cite{hov71};
Hansen \& Travis \cite{ht74}).

    \subsubsection{Scattering matrix}
       \label{subsubsec:scat_mat}

The solution of the radiative
transfer problem with the doubling and adding method requires a suitable
presentation of the scattering matrix $\cal F$ in terms of angular dependence.
Fig. \ref{fig:coord_plane} displays the geometry of scattering for any particular
point inside the slab. We have an incoming ($\mu_{\rm{in}}$) and outgoing
($\mu_{\rm{out}}$, $\Phi_{\rm{out}}$) beams with the scattering angle between
them $\Theta_{\rm{scat}}$. The theory of single light scattering provides
the phase matrix $\cal F$ as a function of scattering angle
[e.g., see (\ref{eq:P}), (\ref{eq:P_Mie}), (\ref{eq:P_Ray}) and
(\ref{eq:P_ave})], where here the scattering plane is the natural plane of
reference. However, both incoming and outgoing beams have their own reference
planes, defined by the Z-axis and a respective beam. Thus, we should perform
a few transformations in order to get the scattering matrix $\cal F$ from
the phase matrix $\cal P$ for any particular beam
(Chandrasekhar \cite{chandrasekhar60}):
\begin{equation}
{\cal F}(\mu_{\rm{out}},\Phi_{\rm{out}}, \mu_{\rm{in}} ) = 
    {\cal R}(i_2-\pi) \; {\cal P}(\cos \Theta_{\rm{scat}}) \;
    {\cal R}(i_1),
      \label{eq:F}
\end{equation}
where $\cal R$ is the polarization rotation matrix, which is a function of
the rotation angle $i$:
\begin{equation}
{\cal R}(i) = \left[ \begin{array}{cccc}
            1 & 0       & 0        & 0\\
            0 & \cos 2i & -\sin 2i & 0\\
            0 & \sin 2i &  \cos 2i & 0\\
            0 & 0       & 0        & 1
            \end{array} \right],
      \label{eq:R}
\end{equation}
Note that this form of $\cal R$ is valid for the intensity vector
representation $(I,Q,U,V)$, adopted by us.
In (\ref{eq:F}), the matrix $\cal R$ first transforms the intensity vector
of the incident beam from its reference plane $A^{\rm{in}}OZ$
into the scattering plane $A^{\rm{in}}OA^{\rm{out}}$ by rotating through
the angle $i_1$ between the two planes (see Fig. \ref{fig:coord_plane}).
The matrix $\cal P$ then gives the intensity of the emergent beam,
scattered by $\Theta_{\rm{scat}}$. Finally,
the matrix $\cal R$ transforms the intensity of the emergent beam
from the scattering plane into the reference plane of the beam $ZOA^{\rm{out}}$
by rotating through the angle $i_2-\pi$. Note that the transformations
(\ref{eq:F}) affect the linear polarization parameters $Q$ and $U$ only,
leaving the mean intensity $I$ and circular polarization parameter $V$
unchanged. Only the position angle $\theta_{\rm{p}}$
of the polarization plane is changed, while the degree
of linear polarization $p_{\rm{l}}=\sqrt{Q^2+U^2}/I$ is unchanged.
The scattering angle $\Theta_{\rm{scat}}$, and the angles $i_1$
and $i_2$ are calculated from the angles $\mu_{\rm{in}}$,
$\mu_{\rm{out}}$ and $\Phi_{\rm{out}}$:
\begin{equation}
\cos \Theta_{\rm{scat}} =
      \mu_{\rm{in}} \mu_{\rm{out}} +
      \sqrt{(1-\mu_{\rm{in}}^2)(1-\mu_{\rm{out}}^2)} \cos \Phi_{\rm{out}}
      \label{eq:Theta_scat}
\end{equation}
\begin{equation}
\cos i_1 = {
           { \mu_{\rm{out}} \sqrt{1-\mu_{\rm{in}}^2} -
             \mu_{\rm{in}} \sqrt{1-\mu_{\rm{out}}^2} \cos \Phi_{\rm{out}}
           }
          \over
           {
             \sin \Theta_{\rm{scat}}
           }
         }
      \label{eq:i1}
\end{equation}
\begin{equation}
\cos i_2 = {
           { \mu_{\rm{in}} \sqrt{1-\mu_{\rm{out}}^2} -
             \mu_{\rm{out}} \sqrt{1-\mu_{\rm{in}}^2} \cos \Phi_{\rm{out}}
           }
          \over
           {
             \sin \Theta_{\rm{scat}}
           }
         }
      \label{eq:i2}
\end{equation}

For computational purposes, it is useful to expand the scattering
matrix in a Fourier series in azimuth angle $\Phi_{\rm{out}}$:
\begin{equation}
{\cal F}(\mu_{\rm{out}},\Phi_{\rm{out}}, \mu_{\rm{in}}) =
  \sum_{m=0}^{\infty}  \left[
    {\cal F}_m^{\rm{c}}(\mu_{\rm{out}},\mu_{\rm{in}}) \; \cos m\Phi_{\rm{out}} +
    {\cal F}_m^{\rm{s}}(\mu_{\rm{out}},\mu_{\rm{in}}) \; \sin m\Phi_{\rm{out}}
    \right]
      \label{eq:F_fou1}
\end{equation}
Then the Fourier modes ${\cal F}_m^{\rm{c}}$ and ${\cal F}_m^{\rm{s}}$
can be calculated with a computationally efficient fast Fourier
transform (FFT) for each pair of $\mu_{\rm{in}}$ and $\mu_{\rm{out}}$:
\begin{equation}
{\cal F}_m^{\rm{c}}(\mu_{\rm{out}},\mu_{\rm{in}}) + i
{\cal F}_m^{\rm{s}}(\mu_{\rm{out}},\mu_{\rm{in}}) =
   {(2-\delta_{m0}) \over N}
       \sum_{n=0}^{N-1}
       e^{{{2\pi} \over {Nnm}}i} \;
        {\cal F}(\mu_{\rm{out}},\Phi_{\rm{out}}^n, \mu_{\rm{in}}),
      \label{eq:F_fou2} 
\end{equation}
where $\Phi_{\rm{out}}^n = 2\pi n/N$ for $n$ ranging from 0 to $N$--1, where
$N$ must be a power of 2.

In this paper we deal with mixtures of spherical grains for which
the phase matrix $\cal P$ has a simple form (\ref{eq:P}).
For this type of $\cal P$, the scattering matrix ${\cal F}$ has the
following symmetries:
1. Changing the sign of $\mu_{\rm{out}}$ and $\mu_{\rm{in}}$ results in a change
of sign of the off-diagonal 2-by-2 blocks of the matrix. Thus, only half the number
of $\mu$s is needed to compute the whole scattering matrix.
2. The upper left and lower right 2-by-2 blocks of ${\cal F}$ are even 
functions of $\Phi_{\rm{out}}$, while the upper right and lower left blocks
are odd functions. Thus, the scattering matrix for $\pi < \Phi_{\rm{out}} \le 2\pi$
can be computed from the values for $0 < \Phi_{\rm{out}} \le \pi $.

From the last symmetry it follows that the cosine matrices
${\cal F}_m^{\rm{c}}$ (sine matrices ${\cal F}_m^{\rm{s}}$)
have off-diagonal (diagonal) blocks of zeros. This means that
we can compose a {\em single} 4-by-4 scattering matrix for each azimuth mode
and pair of zenith angles. Defining a Fourier intensity vector
of the form ${\bar {\bf I}} = (I^{\rm{c}},Q^{\rm{c}},U^{\rm{s}},V^{\rm{s}})$,
containing the cosine azimuth modes of the $I$ and $Q$ Stokes parameters
and the sine modes of $U$ and $V$, then the Fourier scattering matrix
can be represented through the Fourier terms from the FFT as
\begin{equation}
{\cal F}_m(\mu_{\rm{out}},\mu_{\rm{in}}) = \left[ \begin{array}{rrrr}
            {\cal F}_{11}^{\rm{c}} &  {\cal F}_{12}^{\rm{c}} & {\cal F}_{13}^{\rm{s}} & {\cal F}_{14}^{\rm{s}}\\
            {\cal F}_{21}^{\rm{c}} &  {\cal F}_{22}^{\rm{c}} & {\cal F}_{23}^{\rm{s}} & {\cal F}_{24}^{\rm{s}}\\
           -{\cal F}_{31}^{\rm{s}} & -{\cal F}_{32}^{\rm{s}} & {\cal F}_{33}^{\rm{c}} & {\cal F}_{34}^{\rm{c}}\\
           -{\cal F}_{41}^{\rm{s}} & -{\cal F}_{42}^{\rm{s}} & {\cal F}_{43}^{\rm{c}} & {\cal F}_{44}^{\rm{c}}
            \end{array} \right].
  \label{eq:Fm}
\end{equation}
This form of scattering matrix is especially suitable for efficient
implementation of the doubling and adding method (see next subsection).

    \subsubsection{The doubling and adding method}
       \label{subsubsec:doubl_add}

Fortunately, there is an approach to solve the radiative transfer problem
(\ref{eq:rad_tran}--\ref{eq:boundary}), which does not require a direct solution
of the complicated integro-differential equation (\ref{eq:rad_tran}).
This approach is based on the general principle of the linearity
of interaction of electromagnetic radiation with a scattering medium.
Thus, the intensity of radiation leaving a slab is linearly
related to the intensity of the radiation incident on the slab:
\begin{equation}
{\bf I}_{\rm{b}}^{\downarrow} = {\cal T}^{\downarrow\downarrow} {\bf I}_{\rm{t}}^{\downarrow}
                        + {\cal R}^{\downarrow\uparrow} {\bf I}_{\rm{b}}^{\uparrow}
                        + {\bf S}^{\downarrow}, \;\;\;
{\bf I}_{\rm{t}}^{\uparrow} = {\cal T}^{\uparrow\uparrow} {\bf I}_{\rm{b}}^{\uparrow}
                        + {\cal R}^{\uparrow\downarrow} {\bf I}_{\rm{t}}^{\downarrow}
                        + {\bf S}^{\uparrow}, \;\;\;
      \label{eq:inter_princ}
\end{equation}
where ${\bf I}_{\rm{t}}^{\downarrow, \uparrow}$ (${\bf I}_{\rm{b}}^{\downarrow, \uparrow}$)
are the intensity vectors on the top (bottom) of the slab,
${\cal T}^{\downarrow\downarrow}$ and ${\cal T}^{\uparrow\uparrow}$
(${\cal R}^{\downarrow\uparrow}$ and ${\cal R}^{\uparrow\downarrow}$)
are 4-by-4 transmission (reflection) matrices, and ${\bf S}^{\downarrow, \uparrow}$
are the source vectors. The arrows denote the downward ($\downarrow$) and 
upward ($\uparrow$) directions of the beams. In (\ref{eq:inter_princ})
we use a short notation, where the product ${\bf J}$=$\cal A$${\bf I}$
of a matrix $\cal A$ and a vector ${\bf I}$  means: 
\begin{equation}
{\bf J}_i(\mu,\Phi) = {1 \over \pi}
            \int_0^1 \!\! \int_0^{2\pi} \left[ \sum_{i'=1}^4
                {\cal A}_{ii'}(\mu,\mu',\Phi-\Phi') \;
                {\bf I}_{i'}(\mu',\Phi') \right]
                \mu' \; {\rm{d}}{\mu'} \; {\rm{d}}{\Phi'},
      \label{eq:AI}
\end{equation}
where the index $i$, indicating the Stokes parameter, runs form 1 to 4.
A similar rule is valid also for a matrix by matrix multiplication,
e.g. for $\cal C$=$\cal A$$\cal B$
\begin{equation}
{\cal C}_{ii'}(\mu,\mu_{\rm{in}},\Phi) = {1 \over \pi}
            \int_0^1 \!\! \int_0^{2\pi} \left[ \sum_{i''=1}^4
                {\cal A}_{ii''}(\mu,\mu',\Phi-\Phi') \;
                {\cal B}_{i''i'}(\mu',\mu_{\rm{in}},\Phi') \right]
                \mu' \; {\rm{d}}{\mu'} \; {\rm{d}}{\Phi'}.
      \label{eq:AB}
\end{equation}

Equations (\ref{eq:inter_princ}) can be transformed into a form
more suitable for computations by the following steps:
1) expand the matrices and vectors in (\ref{eq:inter_princ}) in Fourier
series in azimuth $\Phi$ and choose the optimum Fourier basis that
would result in decoupling of azimuth modes;
2) discretize zenith angles with a numerical quadrature;
3) rearrange the respective vectors and matrices to their normal
forms as one- and two-dimensional objects.
Below we describe how these steps are realized for relations 
(\ref{eq:AI}) and (\ref{eq:AB}).

Expanding the quantities in (\ref{eq:AI}--\ref{eq:AB}) in
Fourier series in $\Phi$ and choosing a representation for the Fourier
components of vectors as $(I^{\rm{c}},Q^{\rm{c}},U^{\rm{s}},V^{\rm{s}})$
and for matrices the one like (\ref{eq:Fm}) (see section
\ref{subsubsec:scat_mat}), decouples the azimuth modes such that 
for each mode $m$:
${\bar {\bf J}}^m$=${\bar {\cal A}}^m$${\bar {\bf I}}^m$ and
${\bar {\cal C}}^m$=${\bar {\cal A}}^m$${\bar {\cal B}}^m$ where
\begin{equation}
{\bar {\bf J}}^m_{i}(\mu) = 2
            \int_0^1 \!\! \left[ \sum_{i'=1}^4
                {\bar {\cal A}}^m_{ii'}(\mu_{\rm{out}},\mu') \;
                {\bar {\bf I}}^m_{i'}(\mu') \right]
                \mu' \; {\rm{d}}{\mu'},
      \label{eq:AI1}
\end{equation}
\begin{equation}
{\bar {\cal C}}^m_{ii'}(\mu,\mu_{\rm{in}}) = 2
            \int_0^1 \!\! \left[ \sum_{i''=1}^4
                {\bar {\cal A}}^m_{ii''}(\mu,\mu') \;
                {\bar {\cal B}}^m_{i''i'}(\mu',\mu_{\rm{in}}) \right]
                \mu' \; {\rm{d}}{\mu'}.
      \label{eq:AB1}
\end{equation}

We now discretize the zenith angles using a numerical quadrature,
for example, the Gaussian formula.
The expressions (\ref{eq:AI1}--\ref{eq:AB1}) are then transformed into 
\begin{equation}
{\bar {\bf J}}^m_{i,j} = 2
            \sum_{j'=1}^{n_{\rm{\mu}}} \; \sum_{i'=1}^4
                \mu_{j'} \; w_{j'} \; {\bar {\cal A}}^m_{ii',jj'} \;
                {\bar {\bf I}}^m_{i',j'},
      \label{eq:AI2}
\end{equation}
\begin{equation}
{\bar {\cal C}}^m_{ii',jj'} = 2
            \sum_{j''=1}^{n_{\rm{\mu}}} \; \sum_{i''=1}^4
                \mu_{j''} \; w_{j''} \; {\bar {\cal A}}^m_{ii'',jj''} \;
                {\bar {\cal B}}^m_{i''i',j''j'},
      \label{eq:AB2}
\end{equation}
where the indices $j$ and $j'$ indicate the quadrature angles $\mu_j$
and weights $w_j$, and run from 1 to $n_{\rm{\mu}}$, the number of quadrature
angles.

Finally, it is useful to rearrange the intensity vector ${\bar {\bf I}}^m_{i,j}$
for the $m$th azimuth mode from a two-dimensional array into
a one-dimensional vector ${\bar I}^m_{k}$, where $k=i+4(j-1)$ is running
from 1 to 4$n_{\mu}$. If we then respectively redefine the matrices as,
say, ${\bar A}^m_{kk'}=2 \mu_{j'} w_{j'} {\bar {\cal A}}^m_{ii',jj'}$ with
$k=i+4(j-1)$ and $k'=i'+4(j'-1)$ instead of ${\bar {\cal A}}^m_{ii',jj'}$,
the expressions for multiplications of vectors and matrices acquire their
natural form:
\begin{equation}
{\bar J}^m_{k} =
            \sum_{k'=1}^{n} {\bar A}^m_{kk'} \; {\bar I}^m_{k'}, \;\;\;
{\bar C}^m_{k,k'} =
            \sum_{k''=1}^{n} {\bar A}^m_{kk''} \; {\bar B}^m_{k''k'},
      \label{eq:AI3}
\end{equation}
for ${\bar J}^m = {\bar A}^m {\bar I}^m$ and ${\bar C}^m = {\bar A}^m {\bar B}^m$,
respectively, where the vectors have a dimension $n=4n_{\rm{\mu}}$.

We may now return to equations (\ref{eq:inter_princ}) which relates
the intensities of the radiations incident onto and emergent from the slab.
With the new vector basis and new notation, we may rewrite equations
(\ref{eq:inter_princ}) for each azimuth mode (index $m$ is omitted
for better clarity) as ordinary vector-matrix relations:
\begin{equation}
{\bar I}_{\rm{b}}^{\downarrow} = {\bar T}^{\downarrow\downarrow} {\bar I}_{\rm{t}}^{\downarrow}
                        + {\bar R}^{\downarrow\uparrow} {\bar I}_{\rm{b}}^{\uparrow}
                        + {\bar S}^{\downarrow}, \;\;\;
{\bar I}_{\rm{t}}^{\uparrow} = {\bar T}^{\uparrow\uparrow} {\bar I}_{\rm{b}}^{\uparrow}
                        + {\bar R}^{\uparrow\downarrow} {\bar I}_{\rm{t}}^{\downarrow}
                        + {\bar S}^{\uparrow}. \;\;\;
      \label{eq:inter_princ2}
\end{equation}
Equations (\ref{eq:inter_princ2}) are written for the whole slab.
Thus, the matrices ${\bar T}$ and ${\bar R}$ and sources
${\bar S}$ for the whole slab provide the solution of our radiative 
transfer problem.
For an optically thin layer $\tau_{\rm{in}} \ll 1$,
in which single scattering prevails, we may write explicit
expressions for the reflection and transmission matrices and
the source vector (e.g. Evans \& Stephens \cite{ev91}):
\begin{equation}
\mid {\bar R}^{\downarrow\uparrow} \mid^m_{kk'}=
   { {{\tilde \omega} \tau_{\rm{in}}} \over {\mu_j}} \; {{1+\delta_{0m}} \over 4} \; w_{j'}
   \mid {\cal F}_m(\mu_j,-\mu_{j'}) \mid_{ii'},
      \label{eq:R1m}
\end{equation}
\begin{equation}
\mid {\bar R}^{\uparrow\downarrow} \mid^m_{kk'}=
   { {{\tilde \omega} \tau_{\rm{in}}} \over {\mu_j}} \; {{1+\delta_{0m}} \over 4} \; w_{j'}
   \mid {\cal F}_m(-\mu_j,\mu_{j'}) \mid_{ii'},
      \label{eq:R2m}
\end{equation}
\begin{equation}
\mid {\bar T}^{\downarrow\downarrow} \mid^m_{kk'}=
   \delta_{ii'}\delta_{jj'} - { {\tau_{\rm{in}}} \over {\mu_j}}
   \left(
      \delta_{ii'}\delta_{jj'} - {\tilde \omega} \; {{1+\delta_{0m}} \over 4} \; w_{j'}
      \mid {\cal F}_m(\mu_j,\mu_{j'}) \mid_{ii'},
   \right)
      \label{eq:T1m}
\end{equation}
\begin{equation}
\mid {\bar T}^{\uparrow\uparrow} \mid^m_{kk'}=
   \delta_{ii'}\delta_{jj'} - { {\tau_{\rm{in}}} \over {\mu_j}}
   \left(
      \delta_{ii'}\delta_{jj'} - {\tilde \omega} \; {{1+\delta_{0m}} \over 4} \; w_{j'}
      \mid {\cal F}_m(-\mu_j,-\mu_{j'}) \mid_{ii'},
   \right)
      \label{eq:T2m}
\end{equation}
\begin{equation}
\mid {\bar S}^{\downarrow} \mid^m_{k}=
   { {\tau_{\rm{in}}} \over {\mu_j}}
   \mid {\bf S}(\mu_j) \mid^m_{i}, \;\;\;
\mid {\bar S}^{\uparrow} \mid^m_{k}=
   { {\tau_{\rm{in}}} \over {\mu_j}}
   \mid {\bf S}(-\mu_j) \mid^m_{i},
      \label{eq:S1m}
\end{equation}
where $m$ is the azimuth mode, $i,i'$ are the Stokes parameter
indices, $j,j'$ are the quadrature angle indices, $k=i+4(j-1)$
and $k'=i'+4(j'-1)$ and $\mid {\bf S}\mid^m$ is the $m$th Fourier
mode of the source term ${\bf S}$ [eq. (\ref{eq:S})] of the radiative
transfer equation.

The next step is to calculate the quantities ${\bar T}$, ${\bar R}$
and ${\bar S}$ for the whole slab starting from those of the
optically thin layer.
Applying the relations (\ref{eq:inter_princ2}) for two adjacent
layers and eliminating the intensities at the common boundary results in
the expressions for ${\bar T}$, ${\bar R}$ and ${\bar S}$ for
the layer combined from the two layers. This is the essence of
{\em the adding method}. It thus allows us to get the properties of
the combined layer through the properties of the top ($\top$)
and the bottom ($\bot$) layers:
\begin{equation}
\begin{array}{ll}
{\bar R}^{\downarrow\uparrow} = {\bar R}^{\downarrow\uparrow}_{\bot} +
      {\bar T}^{\downarrow\downarrow}_{\bot} {\bar M}^{\downarrow\downarrow}
      {\bar R}^{\downarrow\uparrow}_{\top} {\bar T}^{\uparrow\uparrow}_{\bot}, &
{\bar R}^{\uparrow\downarrow} = {\bar R}^{\uparrow\downarrow}_{\top} +
      {\bar T}^{\uparrow\uparrow}_{\top} {\bar M}^{\uparrow\uparrow}
      {\bar R}^{\uparrow\downarrow}_{\bot} {\bar T}^{\downarrow\downarrow}_{\top}, \\
{\bar T}^{\downarrow\downarrow} = {\bar T}^{\downarrow\downarrow}_{\bot} {\bar M}^{\downarrow\downarrow}
      {\bar T}^{\downarrow\downarrow}_{\top}, &
{\bar T}^{\uparrow\uparrow} = {\bar T}^{\uparrow\uparrow}_{\top} {\bar M}^{\uparrow\uparrow}
      {\bar T}^{\uparrow\uparrow}_{\bot}, \\
{\bar S}^{\downarrow} = {\bar S}^{\downarrow}_{\bot} +
      {\bar T}^{\downarrow\downarrow}_{\bot} {\bar M}^{\downarrow\downarrow}
      ({\bar S}^{\downarrow}_{\top} + {\bar R}^{\downarrow\uparrow}_{\top} {\bar S}^{\uparrow}_{\bot}), &
{\bar S}^{\uparrow} = {\bar S}^{\uparrow}_{\top} +
      {\bar T}^{\uparrow\uparrow}_{\top} {\bar M}^{\uparrow\uparrow}
      ({\bar S}^{\uparrow}_{\bot} + {\bar R}^{\uparrow\downarrow}_{\bot} {\bar S}^{\downarrow}_{\top}), \\
{\bar M}^{\downarrow\downarrow} = \left(1-{\bar R}^{\downarrow\uparrow}_{\top}
       {\bar R}^{\uparrow\downarrow}_{\bot}\right)^{-1}, &
{\bar M}^{\uparrow\uparrow} = \left(1-{\bar R}^{\uparrow\downarrow}_{\bot}
       {\bar R}^{\downarrow\uparrow}_{\top}\right)^{-1},
\end{array}
      \label{eq:adding}
\end{equation}
where ${\bar M}^{\downarrow\downarrow, \uparrow\uparrow}$ are the matrices
which represent multiple reflection factors. Physically,
the formulae (\ref{eq:adding}) may be interpreted through the concept
of multiple reflected beams (Hansen \& Travis \cite{ht74}).
The special case of the adding method, when formulae (\ref{eq:adding})
are applied to two {\em identical} layers with the same optical
thickness $\tau$ and dust composition (albedo $\tilde \omega$, and
scattering matrix $\cal F$), is called {\em the doubling method}.

Thus, in order to calculate the matrices ${\bar R}$, ${\bar T}$
and source vectors ${\bar S}$ for a general, inhomogeneous,
and optically thick slab, one needs to: 1. Divide the slab to thinner layers, 
which could be approximated
as homogeneous; the number and thicknesses of the layers are chosen,
depending on the required accuracy.
2. For each homogeneous layer, apply the doubling method,
that is, first calculate the quantities ${\bar R}$, ${\bar T}$ and 
${\bar S}$ for a sufficiently optically thin sublayer, normally
$\tau_{\rm{in}}=10^{-6}-10^{-5}$, with the formulae
(\ref{eq:R1m}--\ref{eq:S1m}). Then build up the quantities
${\bar R}$, ${\bar T}$ and ${\bar S}$ for the whole layer
in a doubling manner using equations (\ref{eq:adding}).
3. Build up the quantities ${\bar R}$, ${\bar T}$ and ${\bar S}$
for the whole slab by reapplying the adding formulae
(\ref{eq:adding}) for all the layers of the slab.

Note that the pseudo-source of the single-scattered radiation
from the central source [equation (\ref{eq:S})] is exponentially
dependent on the optical depth, thus violating the homogeneity
of the layer, required for doubling to work.
In this case, however, it is possible to use Wiscombe's (\cite{wis76})
extension of the doubling method, incorporating this kind of sources.

Once the matrices ${\bar R}$, ${\bar T}$ and sources ${\bar S}$
are computed, we may calculate the Fourier
modes of the intensities of the radiation emergent from the slab,
${\bar I}_{{\rm{b}}[i,j]}^{\downarrow m}$ and
${\bar I}_{{\rm{t}}[i,j]}^{\uparrow m}$,
with formulae (\ref{eq:inter_princ2}). To get the intensities
themselves, we should convert the Fourier modes back into
the azimuth space:
\begin{equation}
\begin{array}{lll}
{\bf I}_{{\rm{b}}[i,j]}^{\downarrow}(\Phi) =
     \sum_{m=0}^{n_m} {\bar I}_{{\rm{b}}[i,j]}^{\downarrow m} \cdot
       \left\{
         \begin{array}{ll}
            \cos m\Phi, & i=1,2\\
            \sin m\Phi, & i=3,4
         \end{array}
       \right\}, &
   i=1,\ldots,4, & j=1,\ldots,n_{\mu},\\
{\bf I}_{{\rm{t}}[i,j]}^{\uparrow}(\Phi) =
     \sum_{m=0}^{n_m} {\bar I}_{{\rm{t}}[i,j]}^{\uparrow m} \cdot
       \left\{
         \begin{array}{ll}
            \cos m\Phi, & i=1,2\\
            \sin m\Phi, & i=3,4
         \end{array}
       \right\}, &
   i=1,\ldots,4, & j=1,\ldots,n_{\mu},\\
\end{array}
     \label{eq:I_az}
\end{equation}
where $i$ is the Stokes parameter index, $j$ is the quadrature zenith
angle index and $n_m$ is the number of azimuth modes.

Using relations (\ref{eq:inter_princ2}), we may derive
expressions for the intensity vectors ${\bar I}^{\downarrow}$ and
${\bar I}^{\uparrow}$ at any depth inside the slab
\begin{equation}
{\bar I}^{\downarrow} = {\bar M}^{\downarrow\downarrow} \left[
         {\bar S}^{\downarrow}_{\top} +
         {\bar T}^{\downarrow\downarrow}_{\top} {\bar I}^{\downarrow}_{\rm{t}} +
         {\bar R}^{\downarrow\uparrow}_{\top} \left(
           {\bar S}^{\uparrow}_{\bot} +
           {\bar T}^{\uparrow\uparrow}_{\bot} {\bar I}^{\uparrow}_{\rm{b}}
         \right)
       \right], \;\;\;
{\bar I}^{\uparrow} = {\bar M}^{\uparrow\uparrow} \left[
         {\bar S}^{\uparrow}_{\bot} +
         {\bar T}^{\uparrow\uparrow}_{\bot} {\bar I}^{\uparrow}_{\rm{b}} +
         {\bar R}^{\uparrow\downarrow}_{\bot} \left(
           {\bar S}^{\downarrow}_{\top} +
           {\bar T}^{\downarrow\downarrow}_{\top} {\bar I}^{\downarrow}_{\rm{t}}
         \right)
       \right],
     \label{eq:I_inside}
\end{equation}
where the matrices ${\bar T}$ and ${\bar R}$ and sources ${\bar S}$
correspond to the layers above ($\top$) and below ($\bot$)
the level, and the matrices ${\bar M}$ are defined through matrices
${\bar R}$ in (\ref{eq:adding}). The Fourier modes of the intensities
${\bar I}^{\downarrow}$ and ${\bar I}^{\uparrow}$ can be converted into
the azimuth space by means of the formulae like (\ref{eq:I_az}).

The polarization properties
of the radiation reflected by the slab are represented by 
the degree of linear polarization, $p^{\rm{t}}_{\rm{lin}}$,
the position angle, ${\theta}^{\rm{t}}_{\rm{p}}$, and the degree of
circular polarization, $p^{\rm{t}}_{\rm{cir}}$,  
\begin{equation}
p^{\rm{t}}_{\rm{lin}} = {
 \sqrt{Q_{\rm{t}}^2 +
       U_{\rm{t}}^2}
  \over
  I_{\rm{t}}
}, \;\;\;
{\theta}^{\rm{t}}_{\rm{p}} = 
  {1 \over 2}
  \arctan {
     U_{\rm{t}}
     \over
     Q_{\rm{t}}
  }, \;\;\;
p^{\rm{t}}_{\rm{cir}} = {
     V_{\rm{t}}
     \over
     I_{\rm{t}}
   },
       \label{eq:pol_slab}
\end{equation}
where ${\bf I}_{{\rm{t}}}^{\uparrow}$=
($I_{\rm{t}}$,$Q_{\rm{t}}$,$U_{\rm{t}}$,$V_{\rm{t}}$),
$\cos 2{\theta}^{\rm{t}}_{\rm{p}}$ should have the same sign as
$Q_{\rm{t}}$, and all the quantities in (\ref{eq:pol_slab}) are written for
the radiation, reflected from the top. Similar expressions may be written
for the transmitted radiation, escaping from the bottom. 

Other useful characteristics of the radiation leaving the slab
are the fluxes of the radiation emergent from top and bottom:
\begin{equation}
\begin{array}{ll}
{\bf F}_{{\rm{t}}[i]}^{\uparrow} =
    \int_0^{2\pi} \!  \int_{0}^1
    {\bf I}_{{\rm{t}}[i]}^{\uparrow}(\mu, \Phi)
       \; {\rm{d}}{\mu} \; {\rm{d}}{\Phi} =
       2\pi \sum_{j=1}^{n_{\mu}}
           w_j \mu_j \; {\bar I}_{{\rm{t}}[i,j]}^{\uparrow 0}, &
    i=1,\ldots,4,\\
{\bf F}_{{\rm{b}}[i]}^{\downarrow} =
    \int_0^{2\pi} \!  \int_{0}^1
    {\bf I}_{{\rm{b}}[i]}^{\downarrow}(\mu, \Phi)
       \; {\rm{d}}{\mu} \; {\rm{d}}{\Phi} =
       2\pi \sum_{j=1}^{n_{\mu}}
           w_j \mu_j \; {\bar I}_{{\rm{b}}[i,j]}^{\downarrow 0}, &
    i=1,\ldots,4,
\end{array}
\end{equation}
where only zero-th azimuth mode of respective intensity need be used.
Only the Stokes parameters $I$ and $Q$ ($i$=1,2) of the fluxes 
${\bf F}_{{\rm{t}}[i]}^{\uparrow}$ and ${\bf F}_{{\rm{b}}[i]}^{\downarrow}$
can be non-zero. The fraction of the incident flux which is
reflected is called the plane albedo:
\begin{equation}
A(\mu_0) = { {{\bf F}_{\rm{t}[0]}^{\uparrow}} \over F_0  },
      \label{eq:alb_plane}
\end{equation}
where $F_0$ is the flux of the incident unpolarized
radiation of the central source, coming at zenith angle $\mu_0$.

  \subsection{Scattering by an optically thick dusty sphere}
       \label{subsec:scat_sphere}

In this section we apply the dusty slab calculations to calculate the
scattering from a dusty sphere. The sphere is assumed to have $\tau\gg 1$,
so that the thickness of the $\tau\sim 1$ surface layer is much smaller
than the radius of the sphere, and thus every surface element can be
treated with the "local slab" approximation. As in the slab calculations,
we assume that the illumination source distance is much larger than the
radius of the sphere, so the incident rays are assumed to be parallel.
The geometry of the scattering is schematically represented
at Fig. {\ref{fig:coord_sphere}}. The radius of the sphere is $R_{\rm{s}}$.
The flux of the incoming unpolarized radiation from the central source
is $F_0$. The coordinate system is chosen in such a way that the $X$ axis
is directed toward the observer, and the direction of the beam incident
from the source is parallel to the $XY$ plane.
The position of any point $S$ on the sphere surface is then defined by
two angles: zenith angle $\theta$ measured from the $Z$ axis and
azimuth $\Phi$ measured from the $X$ axis. 
The angle between the directions from the source and to the observer,
$\Theta_{\rm{obs}}$,
is the same at any point on the sphere surface, and it is equal to
the scattering angle for single scattering.

The angular parameters of the incident and emergent beams,
are shown at Fig. \ref{fig:coord_plane}:
\begin{equation}
\cos \theta_{\rm{in}}(\theta, \Phi, \Theta_{\rm{obs}})=\mu_{\rm{in}}=-\sin \theta \; \cos (\Phi+\Theta_{\rm{obs}}),
   \label{eq:sph_mu_in}
\end{equation}
\begin{equation}
\cos \theta_{\rm{out}}(\theta, \Phi)=\mu_{\rm{out}}=\sin \theta \; \cos \Phi,
   \label{eq:sph_mu_out}
\end{equation}
\begin{equation}
\cos \Phi_{\rm{out}}(\theta, \Phi, \Theta_{\rm{obs}})=
  {
  {\cos \Theta_{\rm{obs}} + \mu_{\rm{in}} \mu_{\rm{out}}}
  \over
  {\sqrt{(1-\mu_{\rm{in}}^2)(1-\mu_{\rm{out}}^2)}}
  }.
   \label{eq:sph_Phi_out}
\end{equation}

The intensity vector ${\bf I}_{\rm{sph}}$ of the radiation scattered by
the sphere is calculated by integrating
the local intensity ${\bf I}^{\rm{loc}}$ over the illuminated part
of the sphere, as seen from the observer direction:
\begin{equation}
{\bf I}_{\rm{sph}}= \int \! \int_S {\bf I}^{\rm{loc}}
    \; \mu_{\rm{out}} \; {\rm d}S,
   \label{eq:sph_intvec}
\end{equation}
where ${\bf I}^{\rm{loc}}$ is the local intensity of the
emergent radiation, as measured in the observer reference plane, 
chosen here to be the scattering plane $A^{\rm{in}}SA^{\rm{out}}$.
${\bf I}^{\rm{loc}}$ is derived from the local emergent intensity,
${\bf I}^{\uparrow}_{\rm{t}}$ [see equation (\ref{eq:I_az})],
defined with respect to the local (meridian) reference plane,
by rotating it with the matrix ${\cal R}$ through a respective angle $\psi$:
\begin{equation}
{\bf I}^{\rm{loc}}={\cal R}(\psi){\bf I}^{\uparrow}_{\rm{t}}.
   \label{eq:Il}
\end{equation}
Substituting into (\ref{eq:sph_intvec}) and (\ref{eq:Il}) the expressions
for the surface element ${\rm d}S=R_{\rm{s}}^2 \sin \theta \; {\rm d} \theta \; {\rm d} \Phi$
and $\mu_{\rm{out}}$ from (\ref{eq:sph_mu_out}), we obtain
\begin{equation}
{\bf I}_{\rm{sph}}(R_{\rm{s}}, \Theta_{\rm{obs}})=
      R_{\rm{s}}^2 \int_{{{\pi} \over {2}} - \Theta_{\rm{obs}} }^{{\pi} \over {2}}
      \! \cos \Phi \; {\rm d}\Phi \;\! \int_{0}^{\pi} \!\! \sin^2 \theta \; {\rm d}\theta
      \cdot {\bf I}^{\rm{loc}}(\theta, \Phi, \Theta_{\rm{obs}}),
   \label{eq:sph_intvec2}
\end{equation}
\begin{equation}
{\bf I}^{\rm{loc}}(\theta, \Phi, \Theta_{\rm{obs}})=
    {\cal R}[\psi(\theta, \Phi)] \cdot
    {\bf I}^{\uparrow}_{\rm{t}}(\theta, \Phi, \Theta_{\rm{obs}})
    ,
    \label{eq:Il2}
\end{equation}
where the integration limits correspond to the illuminated part of the sphere,
as seen by the observer. The intensity ${\bf I}^{\uparrow}_{\rm{t}}$ depends
on $\theta$, $\Phi$ and $\Theta_{\rm{obs}}$ through $\mu_{\rm{in}}$,
$\mu_{\rm{out}}$ and $\Phi_{\rm{out}}$
(eqs \ref{eq:sph_mu_in}--\ref{eq:sph_Phi_out}).
$\psi$ is the rotation angle defined through its cosine and sine:
\begin{equation}
\cos \psi = {{\sin \theta \; \sin \Phi} \over {\sqrt{1-\mu^2_{\rm{out}}}}}, \;\;\;
\sin \psi = {{-\cos \theta} \over {\sqrt{1-\mu^2_{\rm{out}}}}}
    \label{eq:psi}
\end{equation}
It is simple to show that from the specific azimuthal dependences
of various Stokes parameters (see eqs \ref{eq:I_az})
one gets symmetry relations for the intensities at the surface points,
located symmetrically with respect to the equatorial plane $XOY$:
\begin{equation}
\begin{array}{ll}
{\bf I}^{\uparrow}_{{\rm{t}}[i]}(\theta, \Phi, \Theta_{\rm{obs}}) = 
   {\bf I}^{\uparrow}_{{\rm{t}}[i]}(\pi - \theta, \Phi, \Theta_{\rm{obs}}), &
    i=1,2,\\
{\bf I}^{\uparrow}_{{\rm{t}}[i]}(\theta, \Phi, \Theta_{\rm{obs}}) = 
   -{\bf I}^{\uparrow}_{{\rm{t}}[i]}(\pi - \theta, \Phi, \Theta_{\rm{obs}}), &
    i=3,4.
\end{array}
   \label{eq:I_sym}
\end{equation}
Note that these relations are also valid for the intensities
${\bf I}^{\rm{loc}}$. Thus, using relations (\ref{eq:I_sym}) for
the rotated intensities ${\bf I}^{\rm{loc}}$, we may rewrite
the expression (\ref{eq:sph_intvec2}) for each Stokes parameter of
${\bf I}_{\rm{sph}}$ as follows:
\begin{equation}
I_{\rm{sph}}(R_{\rm{s}}, \Theta_{\rm{obs}})=
     2R_{\rm{s}}^2 \int_{{{\pi} \over {2}} - \Theta_{\rm{obs}} }^{{\pi} \over {2}}
      \! \cos \Phi \; {\rm d}\Phi \;\! \int_{0}^{\frac{\pi}{2}} \!\! \sin^2 \theta \; {\rm d}\theta
      \cdot
      {\bf I}^{{\rm{loc}}}_{[0]}(\theta, \Phi, \Theta_{\rm{obs}}),
   \label{eq:Isph}
\end{equation}
\begin{equation}
Q_{\rm{sph}}(R_{\rm{s}}, \Theta_{\rm{obs}})=
     2R_{\rm{s}}^2 \int_{{{\pi} \over {2}} - \Theta_{\rm{obs}} }^{{\pi} \over {2}}
      \! \cos \Phi \; {\rm d}\Phi \;\! \int_{0}^{\frac{\pi}{2}} \!\! \sin^2 \theta \; {\rm d}\theta
      \cdot
      {\bf I}^{{\rm{loc}}}_{[1]}(\theta, \Phi, \Theta_{\rm{obs}}),
   \label{eq:Qsph}
\end{equation}
\begin{equation}
U_{\rm{sph}}(R_{\rm{s}}, \Theta_{\rm{obs}})=
V_{\rm{sph}}(R_{\rm{s}}, \Theta_{\rm{obs}})=0.
   \label{eq:UVsph}
\end{equation}
These symmetry relations imply that the total circular polarization
of light scattered by a sphere is zero (just as there is
no circular polarization from a single scattering by a spherical
grain), and the linear polarization is
\begin{equation}
p_{\rm{sph}}(\Theta_{\rm{obs}}) = { Q_{\rm{sph}} \over I_{\rm{sph}} }.
   \label{eq:p_sph}
\end{equation}
The ratio of the total flux scattered by the sphere
to the incident flux ($F_0 \cdot \pi R^2_{\rm{s}}$) is
the so-called spherical albedo:
\begin{equation}
A_{\rm{sph}} = 2 \! \int_0^1 A(\mu_0) \; \mu_0 \; \rm{d}\mu_0,
   \label{eq:alb_sph}
\end{equation}
where $A(\mu_0)$ is the plane albedo defined by means of 
eq. (\ref{eq:alb_plane}).

Once the local intensity vector ${\bf I}^{{\rm{loc}}}$
($I_{\rm{loc}}$, $Q_{\rm{loc}}$, $U_{\rm{loc}}$, $V_{\rm{loc}}$) is
calculated as a function of the surface coordinates ($\theta$, $\Phi$),
one can build maps of both linear and circular polarization.
For this purpose we define the following parameters:
the local degree of linear polarization $p^{\rm{loc}}_{\rm{l}}$,
the local position angle ${\theta}^{\rm{loc}}_{\rm{p}}$, and
the local degree of circular polarization $p^{\rm{loc}}_{\rm{c}}$
in a usual manner:
\begin{equation}
p^{\rm{loc}}_{\rm{l}}(\theta, \Phi, \Theta_{\rm{obs}}) = {
\sqrt{Q_{\rm{loc}}^2(\theta, \Phi, \Theta_{\rm{obs}}) +
 U_{\rm{loc}}^2(\theta, \Phi, \Theta_{\rm{obs}})}
\over
I_{\rm{loc}}(\theta, \Phi, \Theta_{\rm{obs}})
},
    \label{eq:pl_loc}
\end{equation}
\begin{equation}
{\theta}^{\rm{loc}}_{\rm{p}}(\theta, \Phi, \Theta_{\rm{obs}}) =
    \frac{1}{2} \arctan { U_{\rm{loc}}(\theta, \Phi, \Theta_{\rm{obs}})
        \over Q_{\rm{loc}}(\theta, \Phi, \Theta_{\rm{obs}})},
    \label{eq:thetap_loc}
\end{equation}
\begin{equation}
p^{\rm{loc}}_{\rm{c}}(\theta, \Phi, \Theta_{\rm{obs}}) =
    { V_{\rm{loc}}(\theta, \Phi, \Theta_{\rm{obs}})
        \over I_{\rm{loc}}(\theta, \Phi, \Theta_{\rm{obs}})}.
    \label{eq:pc_loc}
\end{equation}

\section{Results}
       \label{sec:results}

This section describes extensive numerical results obtained with the codes
PRT and PRT-SPH, which are described in detail in the Appendix. The objects we model 
here are optically thick dusty slabs and
spheres, composed of a simple mixture of bare spherical graphite
and silicate grains having a power-law size distribution
$f(a) \sim a^{-3.5}$ for radii 0.005--0.25 $\mu$m. This model was introduced 
by Mathis, Rumpl, \& Nordsieck (\cite{mrn}; hereafter MRN) for interstellar 
medium dust and further developed by
Draine \& Lee (\cite{dl84}). It served as a standard model for interstellar 
dust for many years, though recent new observational data suggest that other
dust models may be more viable (see Zubko \cite{z99} for a short review). 
However, a dust model which is consistent with all the new data is still not
established, and we therefore chose to make our first exploration of the 
effects of optical depth 
and geometrical configuration using the standard MRN model.

  \subsection{Single scattering by MRN dust}
       \label{subsec:sin_scat}

Figure \ref{fig:alb_g} displays the single scattering albedo and
the phase function asymmetry parameter, $g$, of MRN dust 
for the wavelength range 0.05--1 $\mu$m, used throughout this section. 
The optical constants of graphite and `astronomical'
silicate were taken from Laor \& Draine (\cite{ld93}). The albedo
is typically 0.4--0.6, with a significant drop at $\lambda< 2500$~\AA,
whereas $g$ generally increases towards smaller $\lambda$. The 
scattering phase function becomes significantly more forward peaked at 
smaller $\lambda$ due to the increasing size parameter. 
Figure \ref{fig:extin} demonstrates that the extinction curve
of MRN dust reproduces quite well the mean Galactic extinction curve.

Figure \ref{fig:opt_prop_slab} summarizes the scattering characteristics
of optically thin MRN dust as a function of $\lambda$ and as a function of
$\Theta_{\rm{scat}}$. The scattered spectrum is bluer than the incident
spectrum at all $\Theta_{\rm{scat}}$ for $\lambda>2500$~\AA\ 
since the scattering cross section increases with decreasing $\lambda$.
The "bluening" of the scattered flux is most prominent at small  
$\Theta_{\rm{scat}}$ since the scattering gets more forward peaked
with decreasing $\lambda$ (Fig. \ref{fig:alb_g}). The bluening effect
stops below 2500~{\AA} for $\Theta_{\rm{scat}}\ge 30^{\circ}$ due to
the sudden drop in the albedo (Fig. \ref{fig:alb_g}), compounded by
the rising $g$. A rather strong scattering peak due to graphite grains
is predicted at $\lambda\sim 2200$~{\AA} for large
$\Theta_{\rm{scat}}$. Another similar, but much broader peak is expected in
the far UV (700--1000~{\AA}) due to both graphite and silicate grains.

The polarization spectrum $p(\lambda)$ displays a more complicated 
wavelength dependence than the intensity, and it is also strongly
dependent on $\Theta_{\rm{scat}}$. Overall, $p(\lambda)$ tend to decreases
from 1~$\mu m$ to $\sim 2000$~\AA, and then tends to sharply increase to the
far UV. The polarization plane is generally
at right angle to the scattering plane (defined by the incoming and outgoing
rays), excluding a small range of angles at large $\Theta_{\rm{scat}}$
($150^{\circ}-170^{\circ}$), where the polarization rotates
by $90^{\circ}$, but only over a small range of $\lambda$, and
with a very small amplitude. The feature around $\lambda$=0.14 $\mu$m
is also predicted for the extinction curve, but it is not observed
(Fig. \ref{fig:extin}). Thus, this feature is most likely an artifact 
due to imperfect optical constants of silicate, as discussed by
Kim \& Martin (1995).

The angular dependence of the scattered intensity demonstrates
how the scattering becomes strongly forward peaked with decreasing
$\lambda$. At the longest wavelengths the phase function tends towards
the Rayleigh scattering phase function limit. 
The angular dependence of $p$ always shows a clear peak at
$\Theta_{\rm{scat}}$=80--100$^{\circ}$. The shape of this peak is
not always symmetric, and as mentioned above, $p$ rotates by $90^{\circ}$
(indicated by negative values in the plot) for a small range of $\lambda$
at large $\Theta_{\rm{scat}}$.

  \subsection{Optically thick dusty slab}
       \label{subsec:slab}

The scattering characteristics for an optically thick dusty slab depend
on three angles: $\theta_{\rm{in}} (\mu_{\rm{in}})$,
$\theta_{\rm{out}} (\mu_{\rm{out}})$ and 
$\Phi_{\rm{out}}$ (see Figure \ref{fig:coord_plane}),
rather than just $\Theta_{\rm{scat}}$ as in the
optically thin case.

Figure \ref{fig:opt_thick_slab_i} shows the intensity $I_{\rm{t}}$
as a function of $\lambda$ for various values of $\mu_{\rm{in}}$,
$\mu_{\rm{out}}$, and $\Phi_{\rm{out}}$. At a given $\Phi_{\rm{out}}$,
$\mu_{\rm{in}}$, and $\lambda$, $I_{\rm{t}}$ decreases with increasing
$\mu_{\rm{out}}$, i.e. the surface brightness of the slab decreases 
towards face on view. This effect is largest when $\mu_{\rm{in}}$ is
smallest, since then the radiation is incident close to edge on, 
penetrates only a thin layer at the top of the slab, and this layer is 
effectively optically
thin, and of low surface brightness when viewed close to face on. 
The dependence on $\lambda$ is generally weak longward of 2200~\AA,
with a significant drop at $\lambda<$2200~\AA\ due to the albedo
drop.

Figure \ref{fig:opt_thick_slab_ii} shows $p^{\rm{t}}_{\rm{lin}}(\lambda)$
for the same parameters. As shown later, the main physical parameter 
which determines the shape of $p^{\rm{t}}_{\rm{lin}}(\lambda)$ is
$\Theta_{\rm{scat}}$, which is now a function of $\mu_{\rm{in}}$,
$\mu_{\rm{out}}$, and $\Phi_{\rm{out}}$. For example, when $\mu_{\rm{in}}$
is very close to edge on (0.0672), and $\Phi_{\rm{out}}=90^{\circ}$,
then all values of $\mu_{\rm{out}}$ correspond to 
$\Theta_{\rm{scat}}\sim 90^{\circ}$, and all the $p^{\rm{t}}_{\rm{lin}}(\lambda)$
curves in this subpanel nearly overlap.

Figure \ref{fig:opt_thick_slab_iii} shows the position angle
$\theta^{\rm{t}}_{\rm{p}}$ for the same parameters. The position angle is
almost always fixed at right angle to the plane defined by the incoming and
outgoing rays. The plane used to define $\theta^{\rm{t}}_{\rm{p}}$ in the
optically thick case is defined by the slab normal and the outgoing ray, and
thus $\theta^{\rm{t}}_{\rm{p}}$ simply represents the angle between this plane
and the incident/outgoing rays plane. However, this figure is shown since
$\theta^{\rm{t}}_{\rm{p}}$ jumps by $90^{\circ}$ for certain combinations of
parameters. This effect is similar to the rotation seen in the optically thin
case, and occurs here also for large values of $\Theta_{\rm{scat}}$.

Figure \ref{fig:opt_thick_slab_v} shows the circular polarization
$p^{\rm{t}}_{\rm{cir}}$ for the same parameters. The amplitude
of $p^{\rm{t}}_{\rm{cir}}$ is generally well below 1\%, and it can be
both negative and positive. When $\Phi_{\rm{out}}$=0$^{\circ}$ or 180$^{\circ}$,
$p^{\rm{t}}_{\rm{cir}}$=0 for any $\mu_{\rm{in}}$ and $\mu_{\rm{out}}$.

The doubling method does not allow one to separate directly the contribution of 
single and multiple scattering to the total polarization. To explore this
point we plot in Figure \ref{fig:backscat_slab} $p^{\rm{t}}_{\rm{lin}}(\lambda)$
for complete backscattering ($\Theta_{\rm{scat}}=180^{\circ}$), which occurs
for $\mu_{\rm{in}}$=$\mu_{\rm{out}}$ and $\Phi_{\rm{out}}=180^{\circ}$. In this
case single scattering produces no polarization, and the observed polarization
is due to multiple scattering only. The typical values of 
$p^{\rm{t}}_{\rm{lin}}(\lambda)$ are 1--4\%, and the polarization plane is
generally parallel to the slab normal (see right panel of Figure
\ref{fig:coord_plane}). Figure \ref{fig:opt_thick_slab_ii} indicates
that $p^{\rm{t}}_{\rm{lin}}(\lambda)$ is typically 10--20 times larger than 
obtained for multiple scattering, and thus it must be dominated by the
single scattering contribution. The low multiple scattering polarization 
is explained by the fact that the radiation 
field incident on a grain from multiple scattering is much more isotropic
than the primary source illumination. Most of the photons incident on
the grain following multiple scattering move roughly parallel to the slab
surface, and thus they produce a polarization which is generally parallel to
the slab normal.  

Figure \ref{fig:alb_slab} displays the slab albedo $A(\lambda)$.
Its wavelength dependence is very similar to the MRN dust albedo
(Fig. \ref{fig:alb_g}), but the amplitude is significantly lower.
At very small $\mu_{\rm{in}}$ the scattering layer is optically thin
for most $\mu_{\rm{out}}$, and
most photons that scatter upwards escape, while most
photons scattered downward are absorbed, yielding a slab albedo which
is very close to one half the dust albedo. As $\mu_{\rm{in}}$ increases
the scattering layer becomes optically thicker for all $\mu_{\rm{out}}$,
and the escaping fraction decreases. This effect becomes most prominent
at the shortest wavelengths where the scattering becomes most forward
peaked, which decreases the probability the photons will be scattered
back and escape. Thus, the typical dusty slab albedo falls from $\sim 10$\%
in the optical to only a few \% in the far UV.

  \subsection{Optically thick dusty sphere}
       \label{subsec:sphere}

Figures \ref{fig:maps_0.0500}--\ref{fig:maps_1.0000} present maps
of the linear and circular polarization at
$\lambda$=0.05, 0.2124 and 1 $\mu$m for a range of 
$\Theta_{\rm{obs}}$. The illuminated part of the sphere
extends to the terminator at $\Phi_{\rm{term}}=\frac{\pi}{2} - \Theta_{\rm{obs}}$,
which corresponds to a fractional projected illuminated area of
$\delta(\Theta_{\rm{obs}})=\frac{1}{2}(1-\cos\Theta_{\rm{obs}})$.

The polarization at each point on the surface is generally largest at 
$\Theta_{\rm{obs}}\simeq 90^{\circ}$,
and at most angles it is aligned at right angle to the scattering plane
(i.e. along the axis of the sphere), indicating that it is dominated by
single scattering. When $\Theta_{\rm{obs}}$=180$^{\circ}$ the polarization
tends to be aligned radially, indicating it originates in multiple scattering
which produces a polarization aligned along the local surface normal (see
previous section).
The integrated polarization at $\Theta_{\rm{obs}}$=180$^{\circ}$ is obviously
zero, as expected from symmetry. A similar effect occurs at very small
$\Theta_{\rm{obs}}$, where the single scattering polarization amplitude
becomes very small, and the polarization can be dominated by multiple
scattering. For example, when $\Theta_{\rm{obs}}$=2$^{\circ}$ and
$\lambda$=1~$\mu m$ the polarization pattern is radial, and thus dominated by
multiple scattering. But, at $\lambda$=500~{\AA} the polarization 
pattern remains parallel due to the lower dust albedo, and increased forward 
scattering, both of which suppress the effect of multiple scattering. 

The circular polarization maps typically consist of four regions with two 
boundaries across which the polarization changes sign, one is the equatorial
line, $\theta_{\rm{b}}$=0, and the other is the meridional line
$\Phi_{\rm{b}}=\frac{1}{2}(\pi - \Theta_{\rm{obs}})$
(see  Figures \ref{fig:maps_0.0500}--\ref{fig:maps_1.0000}).
There is a critical angle 
$\Theta_{\rm{obs}}^{\rm{crit}}\simeq 60^{\circ}-80^{\circ}$
at which the circular polarization changes sign at all positions, and there
is also a critical wavelength $\lambda^{\rm{crit}}\simeq$700~{\AA}, at which
a similar sign change occurs.
The local amplitude of $p_{\rm{c}}$ reaches a maximum
of around 1--2 \% at 
$\Theta_{\rm{obs}}$=110$^{\circ}$--130$^{\circ}$.
At $\Theta_{\rm{obs}}$=0$^{\circ}$ or 180$^{\circ}$,
$p_{\rm{c}}$=0 throughout the whole illuminated disk, as expected from symmetry.

Figure \ref{fig:opt_prop_sphere} shows the integrated scattering characteristics
of an optically thick dusty sphere as a function of 
$\lambda$ and $\Theta_{\rm{obs}}$. This figure is analogous to 
Fig. \ref{fig:opt_prop_slab} which shows the scattering characteristics
of optically thin dust. The increase in
$I_{\rm{sph}}(\lambda, \Theta_{\rm{obs}})$
with increasing $\Theta_{\rm{obs}}$ is due to the increasing illuminated
fraction of the sphere. The increase in
$I_{\rm{sph}}(\lambda, \Theta_{\rm{obs}})$ is smallest at the shortest
$\lambda$ because of the counteracting effect of the scattering phase function,
which enhances the scattering at small $\Theta_{\rm{obs}}$. Note that the
feature related to the 2200~\AA\ extinction bump changes its nature from an
emission peak at $\lambda\sim 2500$~\AA\ at large $\Theta_{\rm{obs}}$,
to a weak absorption trough at 
$\lambda\sim 2100$~\AA\ at small $\Theta_{\rm{obs}}$. The wavelength
dependence of $p_{\rm{sph}}$ looks quite similar to the one for optically
thin dust, as further discussed below.

The spherical albedo as a function of $\lambda$ is displayed
in Figure \ref{fig:alb_slab}. It is generally rather low, and drops from
$\sim 11$\% in the near IR to $\sim 4$\% in the far UV. 
The spherical albedo is very similar to the albedo of a slab
illuminated at $\mu_{\rm{in}}=0.5$, which is the mean value of $\mu_{\rm{in}}$
for a sphere.

Figure \ref{fig:sphere_slab1} provides a direct comparison of scattering from 
optically thin dust and from an optically thick sphere, where in both
cases the observed spectrum depends on a single angle only
($\Theta_{\rm{scat}}$, or $\Theta_{\rm{obs}}$). The spectrum of
the scattered light is quite different in the two cases. It is significantly
bluened in the optically thin case at $\lambda >$ 2200~{\AA}, and is generally
grey scattered in the optically thick case. Optically thin scattering 
produces a broad deep centered at $\lambda\sim 1500$~\AA, while the optically
thick dust scattering shows an overall drop at $\lambda < 2200$~\AA, with
only a mild rise towards 1000~\AA. In sharp contrast to the intensity dependence
on optical depth, $p(\lambda)$ is remarkably similar for both cases. The
shape of $p(\lambda)$ is almost identical for both cases at a given 
$\Theta$, and only the amplitudes of $p(\lambda)$ differ by $\sim 10$\%.
This similarity reflects the overall small contribution of multiple
scattering to the polarization. The $\Theta$ dependence of $p(\lambda)$ is
also very similar for the optically thin and thick cases. Thus, $I(\lambda)$
serves as a sensitive probe of the optical depth of the scattering
dust, while $p(\lambda)$  serves as an optical depth independent
probe of the dust properties and scattering geometry.

\section{Conclusions}
       \label{sec:conc}

We present extensive numerical results on the scattering and polarization
properties of dust in various configurations. The solution of the
polarized radiative transfer problem is carried out using the adding-doubling 
method, which applies to a plane parallel configuration. This method can be
extended to other optically thick geometries by an appropriate 
integration over their surface. The numerical implementation is made with a set
of codes which take as input a specified set of dielectric functions,
calculates the scattering phase matrix for a spherical grain of a given
size and composition, and integrates the matrix elements over a specified
dust model to obtain the dust scattering phase matrix. The dust scattering phase
matrix is the input for the numerical code which implements the doubling-adding
method, and outputs the Stokes vectors for a large grid of incident and
outgoing rays. This grid provides the input for the final code which integrates
the slab results over the surface of a sphere and outputs the Stokes vector
as a function of scattering angle from the sphere. 

The accuracy of the radiative transfer calculations was verified by comparison
to analytic results available for a Rayleigh scattering atmosphere, and by
comparison to earlier calculations for an MRN dusty sphere. We also used our
code to obtain a simple analytic approximation for the flux transmitted through
an electron scattering slab, which is accurate to 4\% at 
all $\tau$.

In this paper we used this set of codes to explore the effects of optical
depth,  scattering angles, and dust configuration (slab or sphere), on
the scattering and polarization properties of MRN dust from the near IR
(1$~\mu m$) to the far UV (500~\AA). We find that the scattered spectrum 
provides a sensitive probe for the optical depth of the scattering medium,
while the polarization spectrum provides a probe of the dust model and
scattering geometry, which is practically independent of optical depth.
We provide maps of the linear and circular  polarization for scattering
from a sphere which may be useful for interpreting imaging spectropolarimetry
of scattered light in a variety of Galactic and extragalactic objects from 
the near IR to the far UV. 

The MRN model assumes spherical grains, but the fact that extinction in the
Galaxy induces polarization clearly indicates that this assumption is not valid.
How does the grains non sphericity and alignment affects the scattering polarization?
An accurate answer requires detailed calculations which are beyond the scope
of this paper. However, we note that the amplitude of polarization induced
by transmission through an optical depth of unity ($\le 3$\%) 
is about an order of magnitude smaller than the polarization induced by
scattering. Thus, it is plausible to assume that the observed level of
non sphericity and alignment will not have a major effect on the results 
presented here.

In a following paper we explore the polarization signature of various dust models
over a range of possible compositions and grain size distributions. We also
plan to extend our code to handle other plausible geometries, such as a
dusty disk and a dusty torus. We plan to make these codes
publicly available, which will provide a useful tool for interpreting
available spectropolarimetry results, and in particular, help justify new 
observations 
that can teach us about the properties of dust in various Galactic and
extragalactic environments.

\begin{acknowledgments}
This research was funded by a grant from Israel Science Foundation. We thank
the referee John Mathis for pointing out a couple of errors, and for many 
helpful comments.
\end{acknowledgments}

\clearpage

\appendix

\section{Numerical implementation}
       \label{sec:num_impl}

A package of programs was developed to implement the calculations
described above. The core of the package is
the code PRT which calculates the polarized radiative transfer
in a plane-parallel dusty slab using the adding-doubling method.
The light scattering by an optically thick dusty sphere is calculated
with the code PRT-SPH, using the results,
obtained from PRT. 
The codes were all written in the standard ANSI C language, which
allows a very efficient use of memory and CPU time.
In particular, the memory for large multidimensional arrays, like the
reflection and transmission matrices, which are a function of the number of 
Gaussian quadrature angles and the number of azimuth modes, 
is allocated dynamically (i.e. in run time). The codes may be ported to 
other computers where an ANSI C compiler is installed.

\subsection{The code PRT}
       \label{sec:PRT}

The development of PRT was simplified by the availability
of the FORTRAN code RT3 generously provided by K.F. Evans and G.L. Stephens
(Evans \& Stephens \cite{ev91}), which solves the polarized
radiative transfer problem in the plane-parallel case. 
Some useful ideas and algorithms were adapted from RT3, although
the realization of our code is completely different from that of RT3
both in the general structure and in the details of the code. RT3 was also
very valuable for checking PRT during various steps of
the its development.

The formulation of the problem that PRT solves is as follows.
A plane-parallel slab is illuminated by unidirectional
radiation from a point-like source and/or diffuse radiation from both
above and/or below. The slab may be stratified, i.e. composed of
a number of layers, each of which may be treated as homogeneous,
having a specific geometrical thickness and dust grain composition.
The dust in each layer may contain a number of size-distributed
components, which may be spherical grains of various types:
homogeneous, concentric multilayer and/or composite. The main task
of PRT is to compute the angular-dependent
intensity vectors (Stokes vectors) both for the radiation emergent from
the upper and lower surface of the slab (external problem) and for the radiation 
at various levels inside the slab (internal problem). PRT solves this problem
by the following algorithm:
\begin{enumerate}
\item
  reads input (see below), and allocates memory for arrays;
\item
  calculates quadrature angles and weights;
\item
  calculates the scattering matrix for each azimuth mode and
  each layer;
\item
  loop over the azimuth modes:
  \begin{enumerate}
     \item
     loop over the layers:
     \begin{enumerate}
        \item
        initializes the doubling by calculating the local reflection and
        transmission matrices and source vectors for an initial very thin
        sublayer;
        \item
        doubling: computes the matrices and sources for the whole layer;
     \end{enumerate}
     \item
     adding: calculates the matrices and sources for the whole slab;
     \item
     calculates the intensity vectors, both external and internal;
  \end{enumerate}
\item
   converts the intensity vectors from Fourier modes into azimuth space;
\item
   calculates and outputs fluxes of emergent radiation;
\end{enumerate}

The input includes one main 
file and a few additional smaller (one per layer) files. 
A typical main input file
contains the following parameters: 
1. the number of quadrature angles, $n_{\mu}$, normally 16--24;
2. the number of azimuth modes, $n_m$, normally 4--24, but may be larger
           if the phase function is strongly forward peaked;
3. the number of Stokes parameters, $n_{\rm{st}}$, 1--4;
4. quadrature scheme: Gaussian or double Gaussian;
5. the flux from the point source, $F_0$, usually set to 1, and its
zenith angle, $\theta_0$;
6. the temperature of the diffuse radiation, incoming from above
   and below,
           usually set to 0 K;
7. the optical thickness of the initial doubling layer,
           $\tau_{\rm{in}}$, usually 10$^{-6}$--10$^{-5}$;
8. the wavelength, $\lambda$;
9. the number of azimuths, $n_{\Phi}$, in the range 0$^{\circ}$ to
           180$^{\circ}$, for which output intensity vectors are to be
           calculated;
10. for each layer: its geometrical depth and the name of
           the additional input file with dust mixture properties.

A typical additional input file for a layer contains:
1. the dust mass density;
2. number of dust components;
3. for each dust component: type of grain constitution
           (homogeneous, multilayer, composite), mass fraction
           in per cent, name of a file with the size distribution function,
           the number of grain constituents (one for homogeneous grains),
           and for each grain constituent: name of the constituent and
           its volume fraction in the grain.

The output includes the angular-dependent intensities
and fluxes of the radiation at various depths in the slab.
In the simplest case, only the intensities at the top and bottom are 
printed (reflected and transmitted radiation). The results may be stored
in text or in binary form. The binary form is especially suitable
for use by other codes, e.g. when treating the light scattering
by an optically thick sphere.

It is possible to use the Gaussian or the double Gaussian quadrature
schemes in PRT. However, we have found that in order to produce 
results which are suitable for usage by PRT-SPH (for calculating
light scattering by a sphere) the double Gaussian scheme is more preferable,
as it generates an angle grid which contains very small $\mu$, which is
necessary for PRT-SPH to work properly.

The most important differences between the codes PRT and RT3 are
as follow. (1) The memory for the multidimensional matrices and vectors in
PRT is allocated and freed dynamically, which provides the most
efficient scheme of memory usage, while in RT3 the arrays
are statically allocated. (2) In PRT, the scattering matrix for a specified dust
grain mixture is calculated using the Mie theory for individual grains
summing over the mixtures components and grain-size distributions, 
whereas in RT3 the scattering matrix is computed from its expansion in Legendre 
series
in $\cos \Theta_{\rm{scat}}$. (3) PRT uses a more efficient realization
of the FFT procedure with less than 50 lines of C code, whereas
the respective FORTRAN code in the RT3 contains more than 250 lines.
(4) When performing adding and doubling [eqs (\ref{eq:adding})] and calculating
the intensities of the radiation inside the slab [eqs (\ref{eq:I_inside})],
there is a need to work with inverted matrices. RT3 directly inverts
matrices in these cases. In PRT, we proceed more efficiently.
Since the inverted matrices in eqs (\ref{eq:adding}) and (\ref{eq:I_inside})
are combinations like $A^{-1}B$ or $A^{-1}b$ (where $A$ and $B$
are matrices, $b$ is a vector), it is more computationally suitable
to calculate $A^{-1}B$ or $A^{-1}b$ in the following way: first find the 
$LU$-decomposition of matrix $A$ and then perform a backsubstitution with
the columns of $B$ or $b$ (see e.g. Forsythe, Malcolm, \& Moler \cite{fmm77}).
This is more accurate and saves one matrix multiplication.  
(5) In addition to these differences, PRT
was written with the aim of making it more readable and logically
organized, and thus the general structure of PRT differs from RT3.

The improved computational scheme of PRT allows standard Mie theory 
solutions for grains with size parameters up to $10^6$. Thus, we do not
need to resort to the Rayleigh-Gans or geometrical optics approximations
(Bohren \& Huffman 1983) for any combination of grain sizes and wavelengths. 

The accuracy of the results generated by PRT
depends on the following parameters: the initial optical thickness for doubling
$\tau_{\rm{in}}$, the number of quadrature angles $n_{\mu}$, and
the number of azimuth modes~$n_m$. For a proper choice of $n_{\mu}$,
usually $\ge 16$, and $n_m$, the accuracy is set 
by $\tau_{\rm{in}}$. For example, setting $\tau_{\rm{in}}=10^{-6}$ will
result in 5--6 digit precision.

Typical values of running time and virtual memory, required by PRT
for various values of $n_{\mu}$ and $n_m$ are displayed in Table
\ref{tab:tab1}. These were derived by using a dusty layer of
MRN dust. The results correspond to a calculation per one wavelength,
$\lambda$=0.5 $\mu$m, and per one angle of incoming radiation,
$\theta_0=30^{\circ}$. The optical depth of the layer at $\lambda$=0.5
$\mu$m is $\sim 224$; $\tau_{\rm{in}}$ for doubling was set to 10$^{-6}$.
The calculations have been performed at a DEC Alpha workstation 250$^{4/266}$.
The program was compiled with a C compiler GNU gcc.

\subsection{The code PRT-SPH}
       \label{sec:PRT-SPH}

The code PRT-SPH solves the following problem. A sphere
of radius $R_{\rm{s}}$ is illuminated by unpolarized radiation with
flux $F_0$ from a point source (see Figure \ref{fig:coord_sphere}), where
the angle between the directions to the source and the observer
is $\Theta_{\rm{obs}}$. 
 The code PRT-SPH was designed to solve two main tasks:
first, the computation of the linear and circular polarization maps
that requires calculation of the local intensity vectors over the 
surface of the sphere,
second, the calculation of the integrated characteristics:
the intensity and the degree of linear polarization obtained by
integrating the local intensities over the sphere. The local quantities
are calculated using PRT.

The general algorithm that solves the problem is:
\begin{enumerate}
\item
  discretize the angles $\theta$ and $\Phi$ to define small
  local surface elements;
\item
  calculate local intensities and polarization parameters
  at each of the surface elements (eqs. \ref{eq:sph_mu_in}--\ref{eq:psi} and
  \ref{eq:pl_loc}--\ref{eq:pc_loc});
\item
  calculate the total intensity vector and the degree of
  linear polarization (eqs. \ref{eq:Isph}--\ref{eq:p_sph}).
\end{enumerate}

The azimuth angle $\Phi$ is discretized
uniformly from $-90^{\circ}$ to $90^{\circ}$ with  
$n_{\Phi}$ angles. The zenith angle grid is discretized
with $n_{\theta}$ angles in the range 0--180$^{\circ}$ such that
all local elementary squares have equal surface area
$\Delta S(=2\pi R_{\rm{s}}^2/n_{\Phi} n_{\theta}$). The grid angles $\Phi_k$ and
$\theta_k$ are defined by:
\begin{equation} 
\Phi_k =  90^{\circ} \cdot \left[ \frac{2}{n_{\Phi}} \left(k-\frac{1}{2}\right)
         -1 \right], \;\;\; k=1,\ldots,n_{\Phi},
\end{equation} 
\begin{equation} 
\cos \theta_k = 1 - \frac{2k-1}{n_{\theta}}, \;\;\; k=1,\ldots,n_{\theta}.
\end{equation} 

The results of the computations for a slab
obtained with PRT are output into a binary file, which contains
the Fourier modes of the intensity vector of the reflected radiation
${\bar I}_{\rm{t}}^{\uparrow m}$ for a number of fixed grid angles
$\theta_{\rm{in}}$ and $\theta_{\rm{out}}$.
This provides the input to PRT-SPH.
The local intensity vector ${\bf I}_{\rm{t}}^{\uparrow}$ [in eq. (\ref{eq:Il})]
for any specific angles $\theta_{\rm{in}}$, $\theta_{\rm{out}}$ and
$\Phi_{\rm{out}}$ is calculated as follows: first, by interpolating
the Fourier modes for $\theta_{\rm{in}}$ and $\theta_{\rm{out}}$,
next by computing ${\bf I}_{\rm{t}}^{\uparrow}$ with formula (\ref{eq:I_az})
for the required $\Phi_{\rm{out}}$. For the two-dimensional interpolation over
$\theta_{\rm{in}}$ and $\theta_{\rm{out}}$, PRT-SPH uses bicubic splines. 
The accuracy of this approach for calculating ${\bf I}_{\rm{t}}^{\uparrow}$
generally lies well within 0.1\% when a grid 16$\times$16 for $\mu_{\rm{in}}$ 
and $\mu_{\rm{out}}$ is used for the PRT data.

To compute the integral parameters $I_{\rm{sph}}$ and $Q_{\rm{sph}}$ with
formulae (\ref{eq:Isph}--\ref{eq:Qsph}), the respective integrals are
approximated by sums over the surface elements. For the grid of
40$\times$90 for $\theta$ and $\Phi$ over the sphere, typical in our
calculations, the accuracy of the results is better than 0.01--0.1\%.

\section{Code Testing}

\subsection{Comparison with analytic results for Rayleigh scattering}

To test the accuracy of PRT and PRT-SPH,
we performed calculations for a pure Rayleigh scattering sphere.
This corresponds physically to a Thomson scattering electrons sphere.
This case was carefully studied and there are various analytical
results to compare with, e.g. Chandrasekhar
(\cite{chandrasekhar60}).

We calculated the inclination dependence of the intensity
and the degree of linear polarization for radiation transmitted
through an optically thick slab. We used $n_{\mu}$=36,
$n_m$=4, $\tau_{\rm{in}}$=10$^{-6}$ and $\tau$=1460 for the total optical depth 
of the scattering layer. The calculations made with PRT
agree with the analytical results of Chandrasekhar
(\cite{chandrasekhar60}) to better than 0.05\%.

The optical depth dependence of the flux of radiation transmitted
through a Rayleigh scattering slab is shown in Figure
\ref{fig:rayleigh2}, as calculated with PRT. For comparison, we plot 
the large-$\tau$ asymptotic analytic solution given by
Yanovitskij (1997):
\begin{equation}
F_{\rm{1}}(\tau)={4 \over 3} {1.265 \over {\tau+1.423}}
       \label{eq:F_large_tau}
\end{equation}
We find an agreement to within 1\% between the analytic and numerical
solutions at $\tau>$3.

As a side note, we tried to use our numerical solution, which is exact
at all $\tau$, to obtain an improved analytic approximation which
generalizes equation \ref{eq:F_large_tau} to a larger range of $\tau$.
We found
a surprisingly good approximation to our result which is valid for {\em all}
$\tau$.
\begin{equation}
F_{\rm{2}}(\tau)={4 \over 3} {1.265 \cdot a \over {\tau+1.423 \cdot b}}
       \label{eq:F_all_tau}
\end{equation}
with $a$=1.019 and $b$=1.194. The accuracy of this simple approximation
is 0.3--2\% for $\tau <0.5$, 2--4\% for $0.5<\tau <5$, 
and 0.5--2\% for $5<\tau $.

In a pure scattering atmosphere no flux is lost due to absorption,
and thus energy conservation implies that the incident flux must equal the 
sum of transmitted and
scattered fluxes. The right panel in Figure \ref{fig:rayleigh2}
shows the accuracy to which energy is conserved in PRT. 
There error increases roughly linearly with $\tau$ due
to accumulation of errors in the doubling procedure.
However, the errors for $\tau\sim 100-1000$, typical for this study,
are still very small ($<$0.0001--0.001\%).

\subsection{Comparison with previous results for a dusty sphere}

White (\cite{white2}) calculated the scattering
properties of an optically thick dusty sphere with an MRN composition, also
using a doubling method for the radiative transfer. 
Figure \ref{fig:white1} shows a comparison of the wavelength-dependent spherical 
albedo, maximum linear polarization, and ratio of forward to backward scattering,
taken from the figures in White (\cite{white2}) to those calculated with
PRT and PRT-SPH. The overall consistency is quite satisfactory, considering
the significant improvements in the optical constants for graphite and silicate
grains since the study of White (\cite{white2}).

\section{Electron scattering versus dust scattering sphere}

Figure \ref{fig:rayleigh3} presents maps of the linear polarization
of an optically thick electron scattering
sphere, obtained with PRT-SPH.
Comparison with corresponding maps of dusty spheres (Figs
\ref{fig:maps_0.0500}--\ref{fig:maps_1.0000}) indicates that, as expected,
multiple scattering has a larger contribution to the polarization
in the electron scattering sphere. For example, 
at $\Theta_{\rm{obs}}$=180$^{\circ}$, the local degree
of $p$ reaches 7--8\%, which is twice the value
for a dusty sphere, and a similar effect is seen at very small 
$\Theta_{\rm{obs}}$. On the other hand, the maximum local 
polarization (at $\Theta_{\rm{obs}}$=90$^{\circ}$), is
55--60\% which is significantly lower than the maximum values
of up to 90\% (for $\lambda$=0.05 $\mu$m) found for a dusty 
sphere. Dust absorption enhances the polarization since it
suppresses the contribution of multiple scattering, which adds
unpolarized light, or light which is polarized at right angle
to the single scattered light (for $\Theta_{\rm{obs}}$=90$^{\circ}$),
in both cases lowering the total polarization.

Figure \ref{fig:rayleigh4} presents the integrated scattering characteristics of 
the electron scattering sphere in comparison to those of a dusty sphere.
The electron scattering sphere scatters light more efficiently
at most angles ($\Theta_{\rm{obs}}>$20$^{\circ}$). Its albedo is
essentially unity, versus $4-11$\% only for a dusty sphere.
However, a dusty sphere is a much more efficient polarizer, allowing
maximum $p_{\rm{sph}}$ of 70--80\%, versus only about 30\% for the
electron scattering sphere, again due to the greater contribution
of multiple scattering in the later case.

Note that the integrated electron scattering sphere polarization 
is rotated by 90$^{\circ}$ for $\Theta_{\rm{obs}}$=0-20$^{\circ}$.
This effect is not seen in the Monte Carlo calculations of Code
\& Whitney (1995), most likely due to the very low intensity
of light scattered at this range, coupled with the low polarization
amplitude. This demonstrates the advantage of an accurate solution
over Monte Carlo calculations, though the former is limited to the
$\tau \gg 1$ case only.

\clearpage

\begin{deluxetable}{cccc}
\tablewidth{0pt}
\tablecaption{CPU time and virtual memory, required by PRT.  \label{tab:tab1}}
\tablehead{
  \colhead{$n_{\mu}$} &
  \colhead{$n_m$} &
  \colhead{CPU time} &
  \colhead{memory size}
}
\startdata
4   &   1    &  3$\fs$8 & 2.81 Mb\nl
4   &   8    &  5$\fs$0 & 2.90 Mb \nl
4   &  16    &  6$\fs$7 & 3.01 Mb \nl
4   &  32    &  8$\fs$6 & 3.21 Mb \nl
4   &  64    & 13$\fs$4 & 3.62 Mb \nl
8   &   1    &  6$\fs$5 & 3.05 Mb \nl
8   &   8    & 21$\fs$0 & 3.38 Mb \nl
8   &  16    & 26$\fs$0 & 3.77 Mb \nl
8   &  32    & 48$\fs$0 & 4.50 Mb \nl
8   &  64    &  1$\fm$5 & 6.00 Mb \nl
16  &   1    & 36$\fs$0 & 3.98 Mb \nl
16  &   8    &  2$\fm$2 & 5.24 Mb \nl
16  &  16    &  4$\fm$0 & 6.69 Mb \nl
16  &  32    &  8$\fm$5 & 9.56 Mb \nl
16  &  64    & 15$\fm$2 & 15.3 Mb \nl
32  &   1    &  2$\fm$0 & 5.05 Mb \nl
32  &   8    &  8$\fm$5 & 7.37 Mb \nl
32  &  16    & 16$\fm$0 & 10.0 Mb \nl
32  &  32    & 32$\fm$0 & 15.3 Mb \nl
32  &  64    &  1$\fh$0 & 25.9 Mb \nl
\enddata
\end{deluxetable}

\clearpage

\figcaption[    ]{
Schematic representation of light scattering by a slab of 
optical thickness $\tau_0$.
Unpolarized radiation with a flux $F_0$ is
incident from a zenith angle $\theta_0$.
The downward and upward intensities ($I^{\downarrow}$ and $I^{\uparrow}$)
are a function of $\tau$, $\mu=\cos \theta$ and
azimuth angle $\Phi$. Diffuse radiation may be incident 
from the top ($I^{\downarrow}_{\rm{t}}$) and the bottom
($I^{\uparrow}_{\rm{b}}$). The radiation which emerges from the top
($I^{\uparrow}_{\rm{t}}$) and bottom ($I^{\downarrow}_{\rm{b}}$),
results from the transfer which is expressed by the reflection
($R^{{\downarrow}{\uparrow}}$, $R^{{\uparrow}{\downarrow}}$) and
transmission ($T^{{\downarrow}{\downarrow}}$, $R^{{\uparrow}{\uparrow}}$)
matrices, plus a possible contribution of internal sources ($S^{\downarrow}$,
$S^{\uparrow}$).
\label{fig:coord_slab}}

\figcaption[ ]{
The scattering geometry for a plane-parallel slab.
The slab is in the $XY$ plane and $Z$ is the normal.
The incoming beam makes an angle $\theta_{\rm{in}}$ with
the $Z$ axis, and the outgoing beam is in the direction 
$\theta_{\rm{out}}$, $\Phi_{\rm{out}}$, where $\Phi_{\rm{out}}$ is the azimuth 
difference between
the incoming and outgoing beams.
The scattering angle $\Theta_{\rm{scat}}$ is the angle between the continuation
of the incoming beam $A^{\rm{in}}O$ and the outgoing beam $OA^{\rm{out}}$.
The rotations angles of the incoming (outgoing) beam  $i_1$ 
($i_2$) is measured from the meridian plane of
the incoming beam $A^{\rm{in}}OZ$ (the plane $A^{\rm{out}}OZ$),
specified by $n^{\rm{in}}_{\rm{m}}$ ($n^{\rm{out}}_{\rm{m}}$),  
and the scattering plane $A^{\rm{in}}OA^{\rm{out}}$,
specified by $n^{\rm{in}}_{\rm{s}}$ ($n^{\rm{out}}_{\rm{s}}$).
Right panel: A view of the scattering geometry along the outgoing beam, showing 
the plane of linear 
polarization of single scattered radiation,
and of multiply scattered radiation. The first is generally at right angle
to the plane defined by the incoming and outgoing beams, while the second is
generally at right angle to the plane of the slab.
\label{fig:coord_plane}}

\figcaption[ ]{
 The scattering geometry for a dusty sphere. The direction of the incoming beam 
 is $A^{\rm{in}}S$, and it gets scattered by $\Theta_{\rm{obs}}$ along the
 direction $SA^{\rm{out}}$. The scattering point $S$ coordinates on the
 sphere are zenith angle $\theta$ and azimuthal angle $\Phi$ (measured
 from the $X$ axis), and the local normal is along $n(\theta, \Phi)$.
 The scattering at $S$ is calculated using the local slab approximation,
 as presented in Figure 2.
       \label{fig:coord_sphere}}

\figcaption[]{
The single scattering albedo and the phase function asymmetry
parameter $g$ as a function of wavelength. Note the sharp drop in
albedo below 2200~\AA, and the tendency of scattering to become strongly
forward peaked at short wavelengths.
      \label{fig:alb_g}}

\figcaption[]{
The extinction curve for the MRN dust mixture
compared to the mean Galactic extinction curve for $R_V$=3.1
(Cardelli, Clayton, \& Mathis 1989). Note that the Galactic extinction
at 1000~\AA\ is somewhat uncertain.
      \label{fig:extin}}

\figcaption[]{
The scattering and polarization properties of optically thin MRN dust.
Left panels show the $\lambda$ dependence of the normalized intensity
$I(\lambda, \Theta_{\rm{scat}})/I(1$ ${\mu}m, 90^{\circ})$,
degree of linear polarization, $p(\lambda)$, and polarized intensity
$I\times p$. Right panels show the $\Theta_{\rm{scat}}$ dependence
for the same parameters. The polarization plane is generally at right angle
to the scattering plane, excluding the cases where $p(\lambda)<0$ 
(marked by bold lines and filled points in the lower panels), which refer to 
polarization parallel to the scattering plane.
      \label{fig:opt_prop_slab}}

\figcaption[]{
The logarithm of the total intensity $I_{\rm{t}}(\lambda)$ of light scattered
by an optically thick MRN dusty slab. Each row corresponds to a given
$\mu_{\rm{in}}$, each column to a given $\Phi_{\rm{out}}$, and each subpanel
presents curves for a range of $\mu_{\rm{out}}$. The incident flux is unity.
      \label{fig:opt_thick_slab_i}}

\figcaption[]{
The degree of linear polarization for the same parameters as
in Fig. 7.
      \label{fig:opt_thick_slab_ii}}

\figcaption[]{
The position angle of linear polarization for the same parameters as in Fig. 7.
      \label{fig:opt_thick_slab_iii}}

\figcaption[]{
The degree of circular polarization for the same parameters as in Fig. 7.
Note that due to symmetry the circular polarization is zero at
$\Phi_{\rm{out}}$=0$^{\circ}$ and at 180$^{\circ}$.
      \label{fig:opt_thick_slab_v}}

\figcaption[]{
The linear polarization of the light backscattered
(i.e. $\mu_{\rm{in}}=\mu_{\rm{out}}$, $\Phi_{\rm{out}}=180^{\circ})$
by an optically thick slab as a function of $\lambda$ and $\mu$. Only multiple 
scatterings contribute to $p$ in this case, and its amplitude is generally 
smaller by a factor of 10-20
compared with the single scattering contribution.
      \label{fig:backscat_slab}}

\figcaption[]{
The plane albedo of an optically thick dusty MRN slab
as a function of $\lambda$ and $\mu_{\rm{in}}$. The albedo of an
optically thick sphere is also indicated.
      \label{fig:alb_slab}}

\figcaption[]{
Maps of the linear and circular polarization of light scattered by
an optically thick MRN dusty sphere at
$\lambda =0.05~\mu m$ for various $\Theta_{\rm{obs}}$.
Note the very large differences in the relation between arrow size and
polarization amplitude at different $\Theta_{\rm{obs}}$.
      \label{fig:maps_0.0500}}

\figcaption[]{
As in Figure 13 for $\lambda=0.2124~\mu$m.
      \label{fig:maps_0.2124}}

\figcaption[]{
As in Figure 13  for $\lambda=1.0~\mu$m.
      \label{fig:maps_1.0000}}

\figcaption[]{
The scattering and polarization properties of an optically thick MRN dusty
sphere. All parameters are as shown in Fig. 6, excluding the intensity
which is not normalized here.
      \label{fig:opt_prop_sphere}}

\figcaption[]{
Comparison of scattering by optically thin dust versus an optically thick
dusty sphere. Upper two panels show the $\lambda$ dependence, and the lower
two panels the $\Theta$ dependence. Note the large difference in the 
wavelength dependence of $I(\lambda,\Theta)/I(1$~${\mu}m ,\Theta)$
for the optically thin and thick cases, but the remarkable similarity
of $p(\lambda)$.
      \label{fig:sphere_slab1}}

\figcaption[]{
The flux of the radiation transmitted through a Rayleigh
scattering slab as a function of optical thickness
({\it left panel}) as calculated with PRT.
The incident flux is unity with $\mu_{\rm{in}}$=1.
The large-$\tau$ analytic asymptotic solution
$F$=${4 \over 3} {1.265 \over {\tau+1.423}}$ 
of Yanovitskij (1997), agrees well with our results.
We also show our improved approximation
$F$=${4 \over 3}{1.265 \cdot a \over {\tau+1.423 \cdot b}}$ with $a$=1.019
and $b$=1.194, which is accurate to 4.0\% 
for all $\tau$.
The accumulated error in flux conservation, as calculated with PRT 
for the Rayleigh scattering
atmosphere, is shown in the {\it right panel}. 
      \label{fig:rayleigh2}}

\figcaption[]{Comparison with the earlier calculations of
White (1979b) for the spherical albedo, maximum linear polarization, 
$p_{\rm{max}}$, and ratio of forward to backward scattering, $\rho$, for 
scattering by an optically thick dusty sphere of the MRN dust composition.
The agreement is acceptable, given the updates in the dielectric function
since the study of White.
      \label{fig:white1}}

\figcaption[]{
Maps of linear polarization of the light scattered by an optically
thick Rayleigh scattering sphere (electron scattering) for various
$\Theta_{\rm{obs}}$.
      \label{fig:rayleigh3}}

\figcaption[]{Comparison of the scattering properties of an optically thick
electron scattering sphere versus an optically thick dusty sphere.
The intensity of scattered light is generally much higher for an
electron sphere, but its integrated polarization is significantly
lower.
      \label{fig:rayleigh4}}

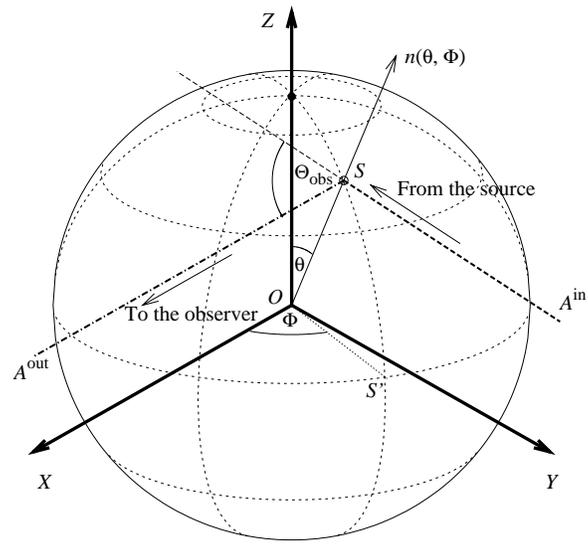
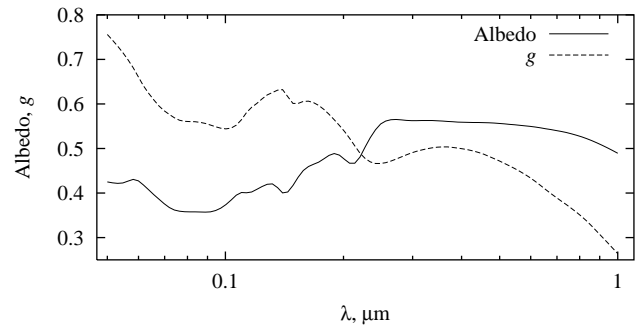
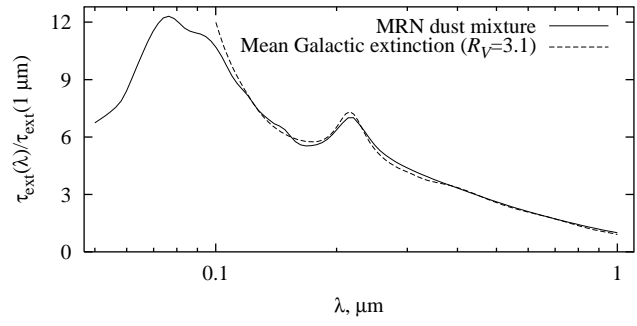
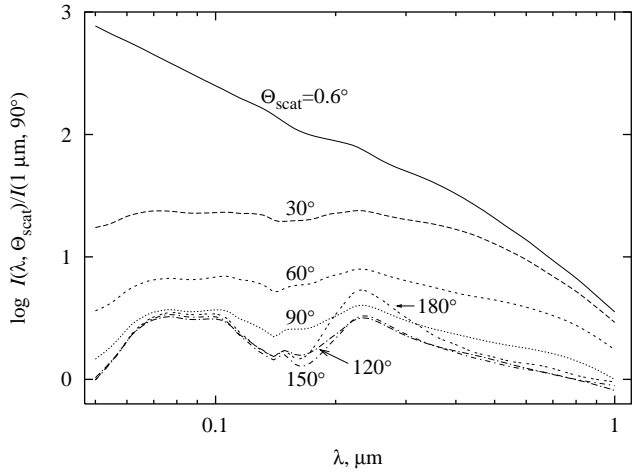
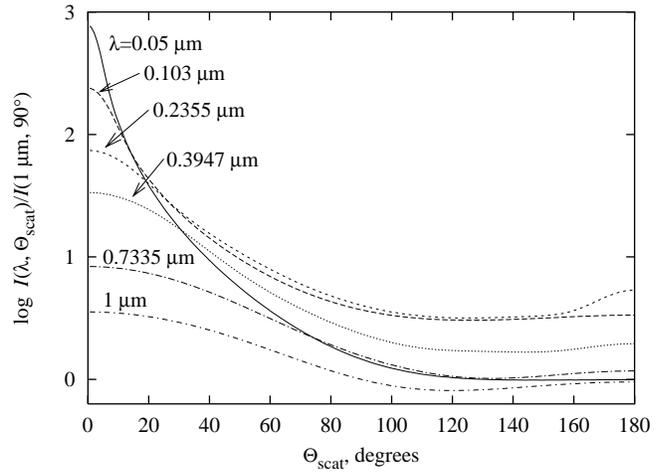
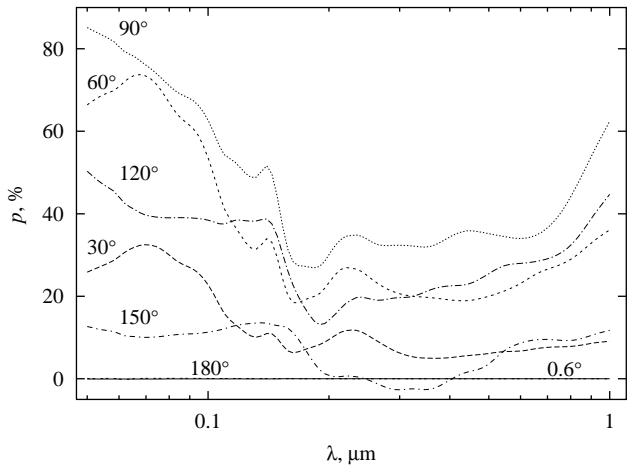
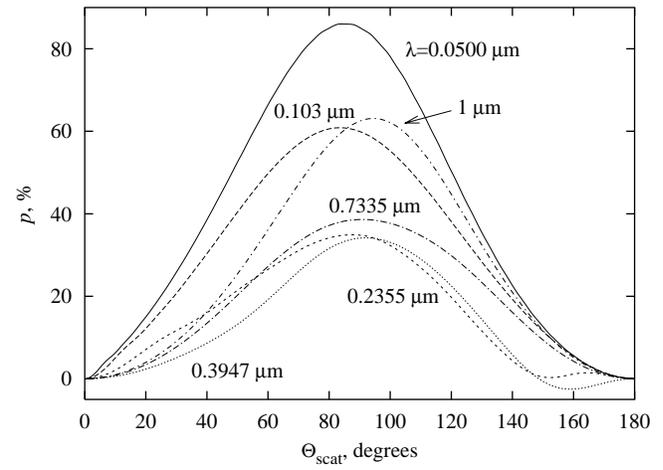
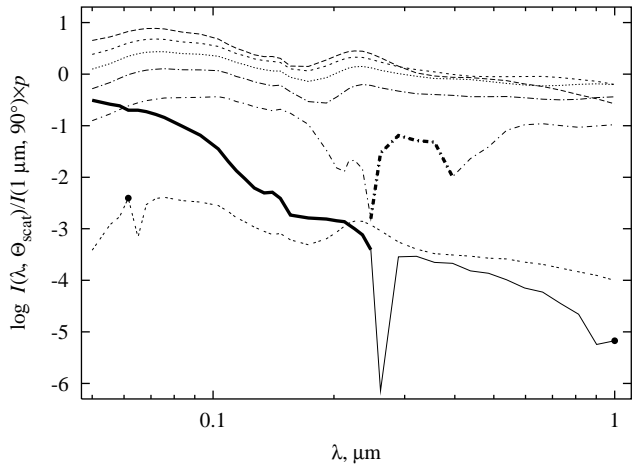
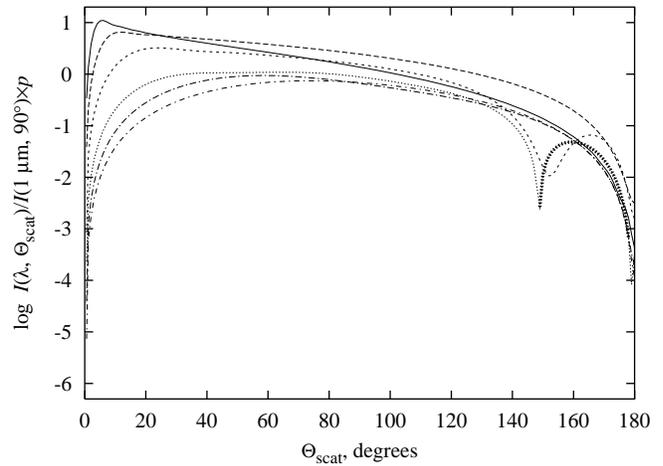
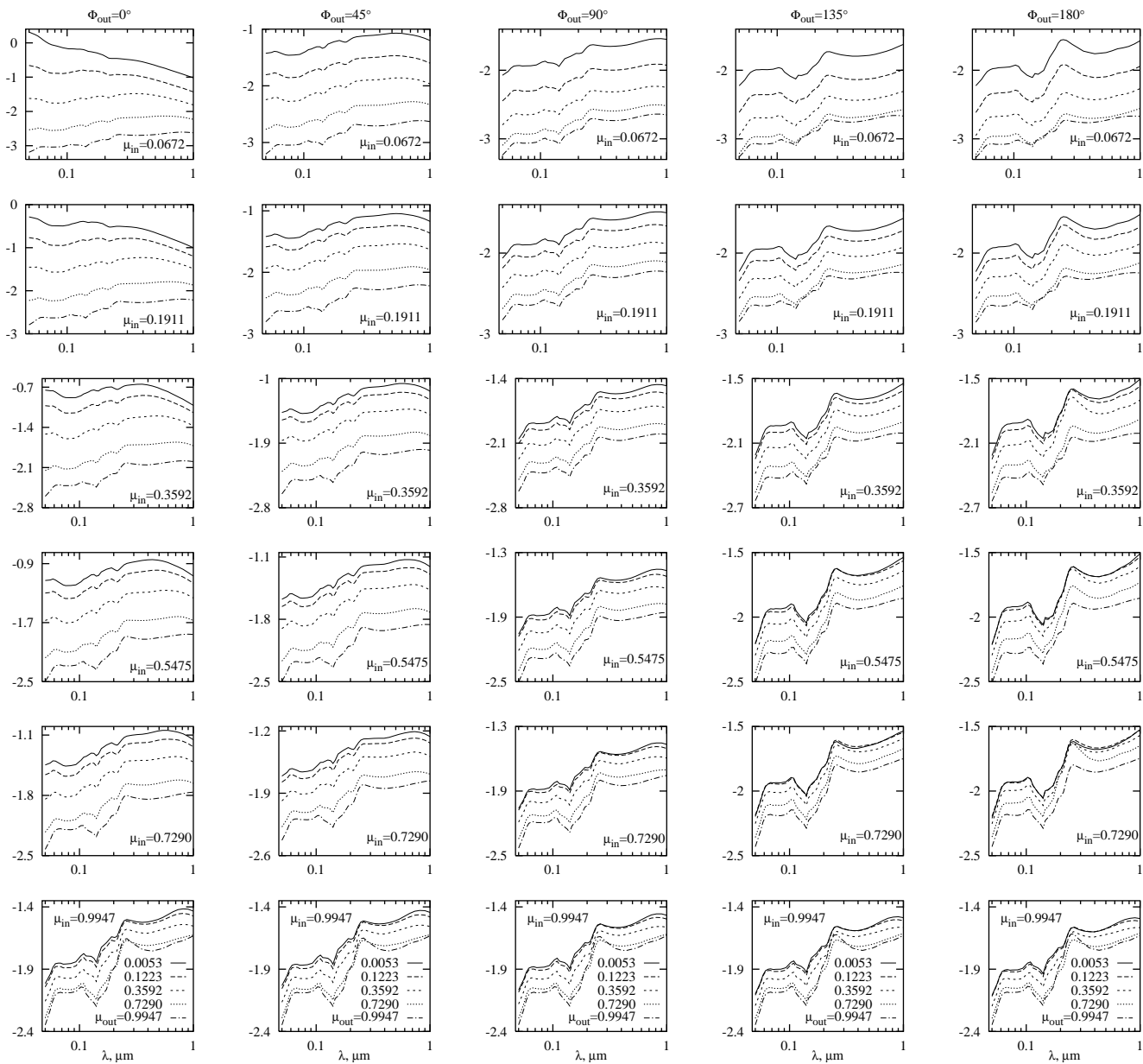
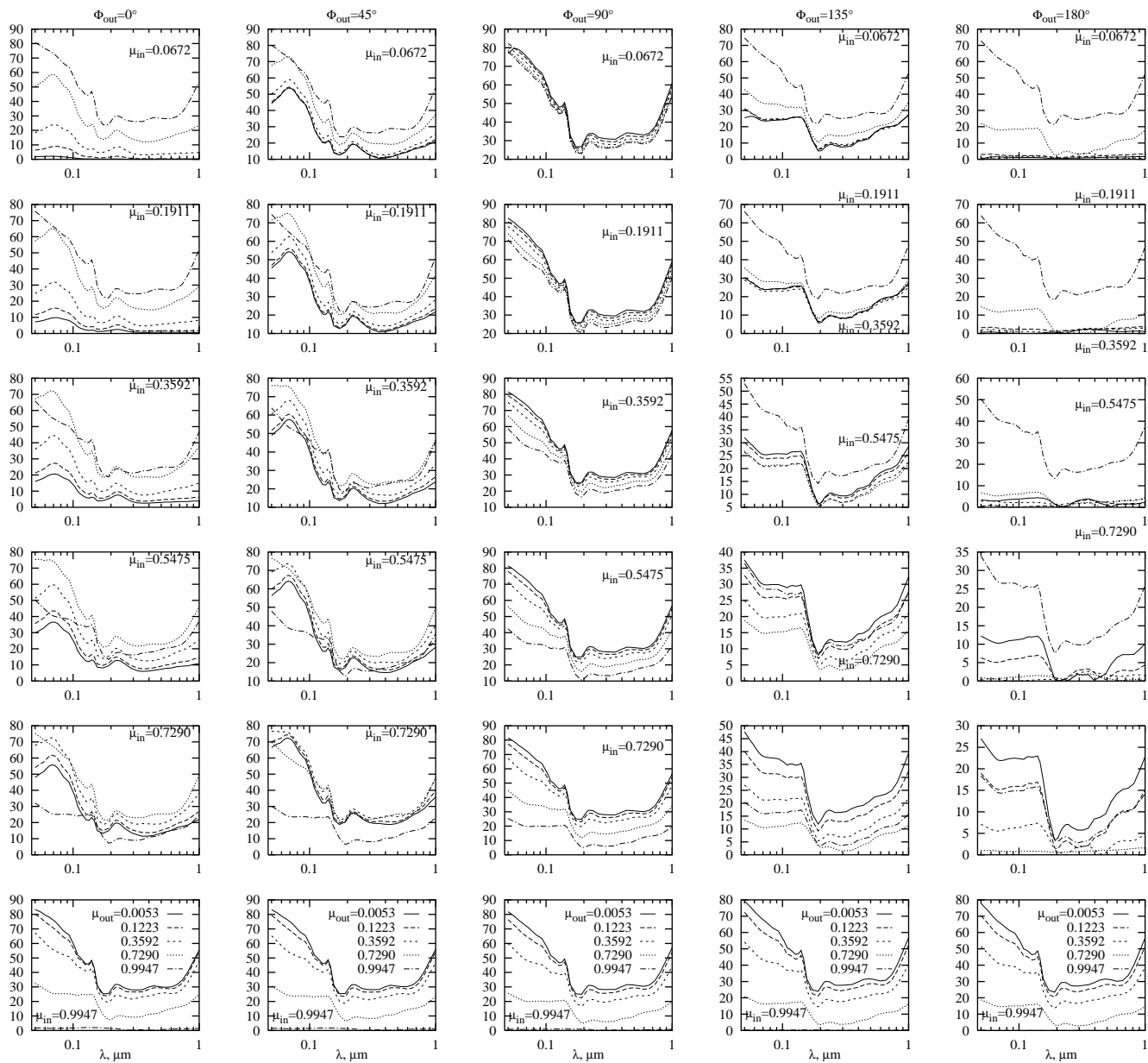
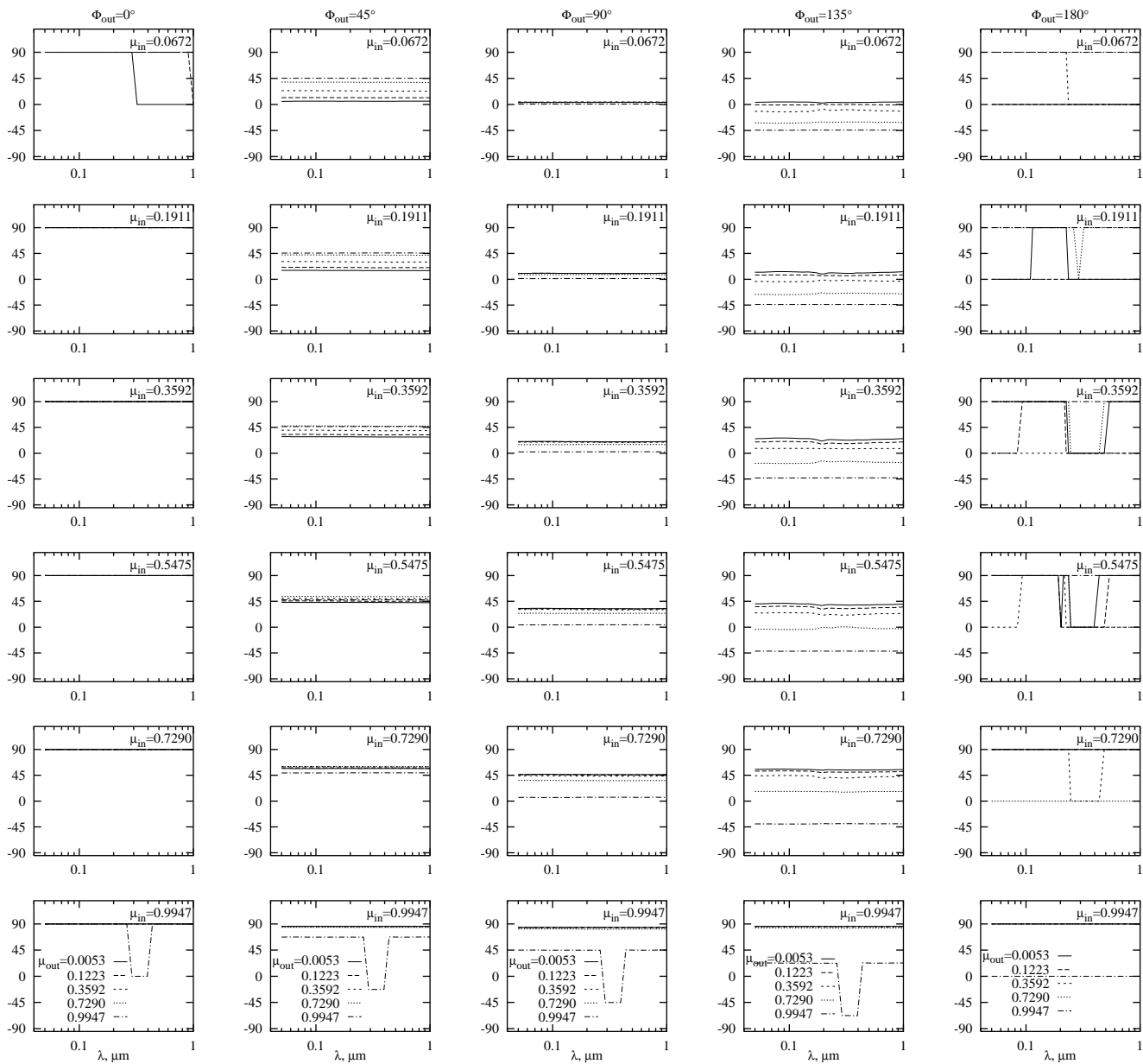
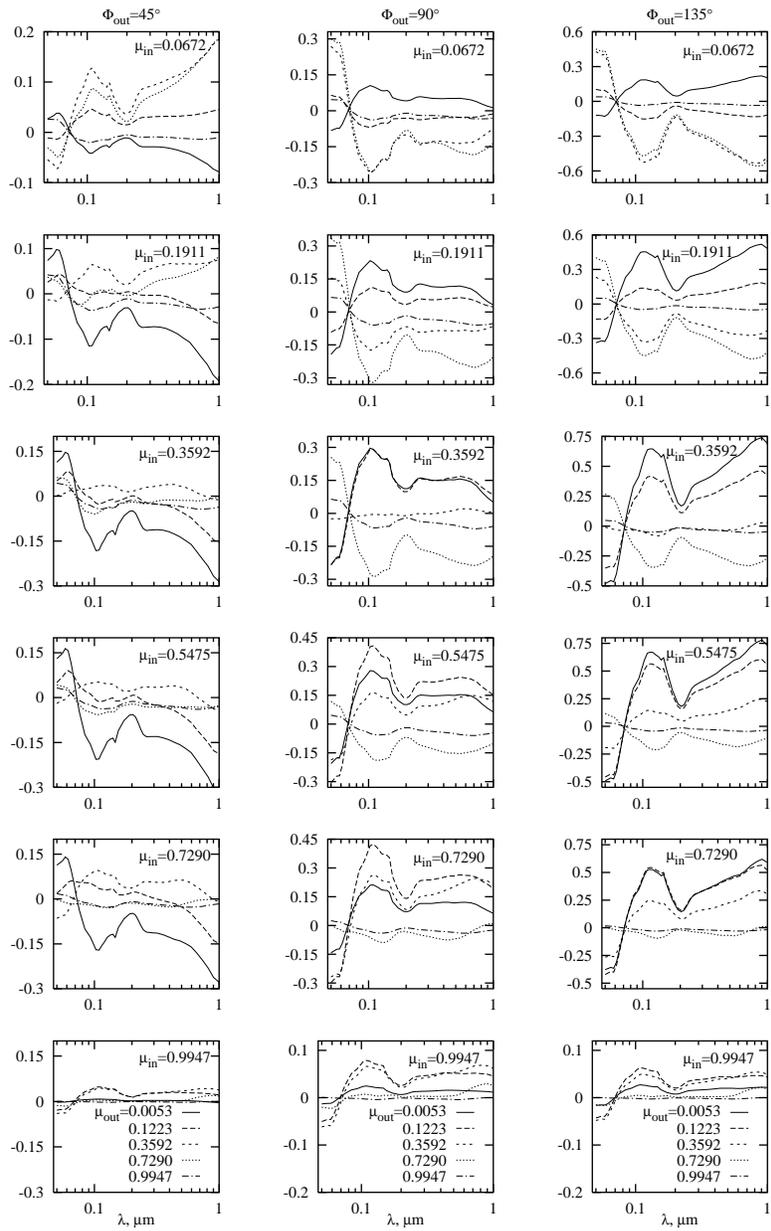
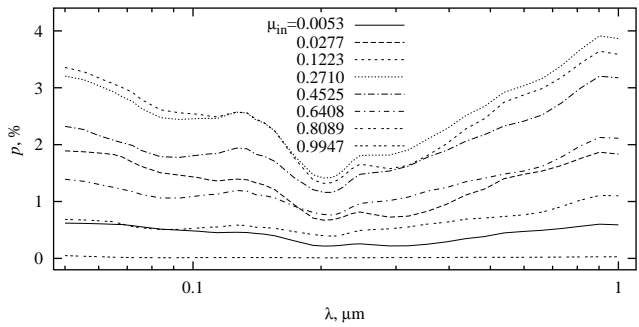
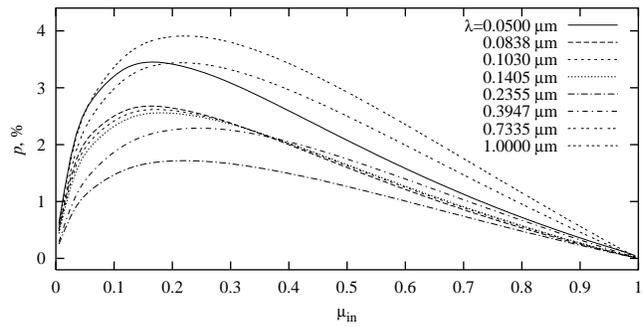
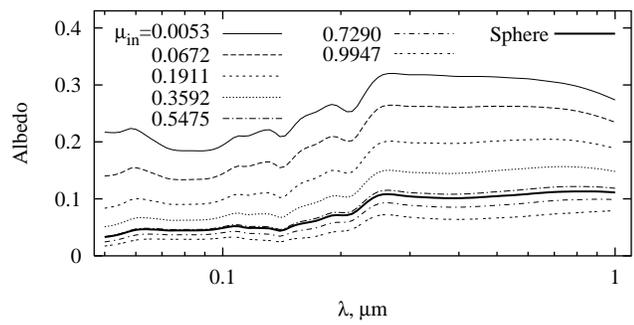
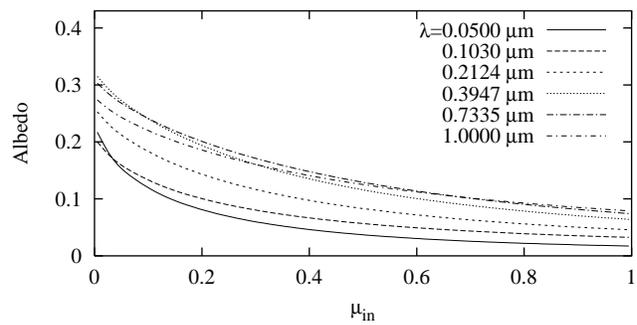
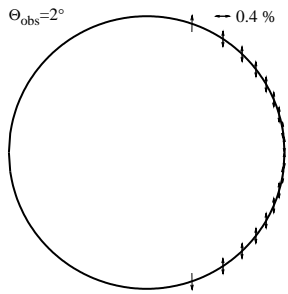
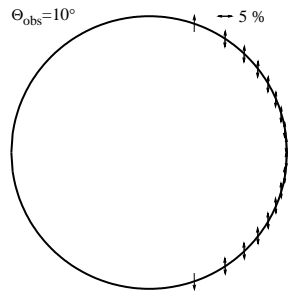
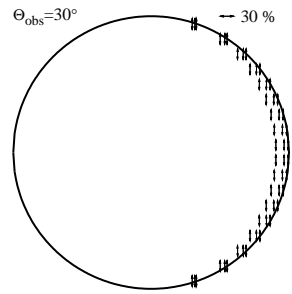
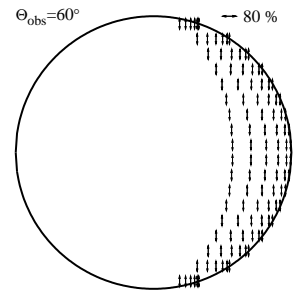
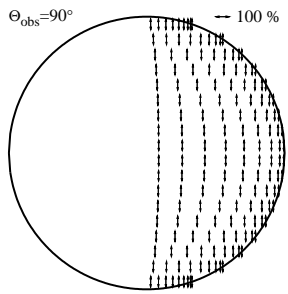
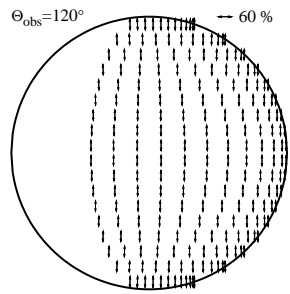
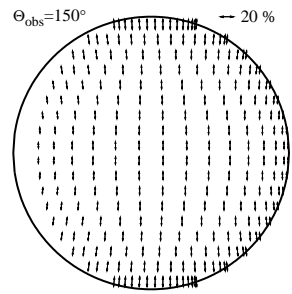
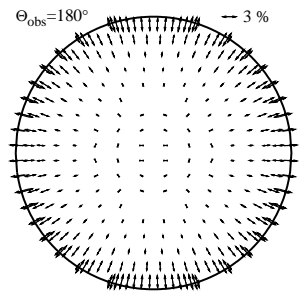
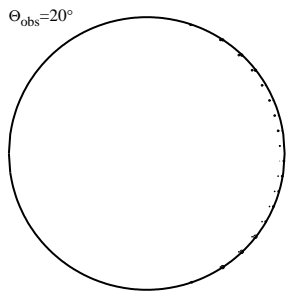
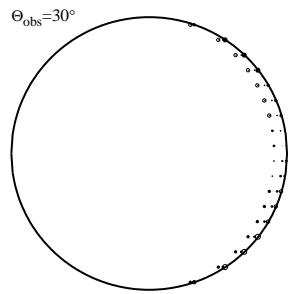
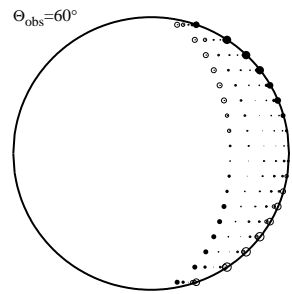
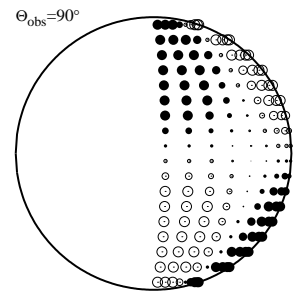
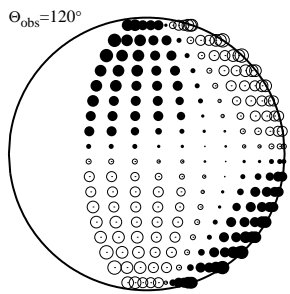
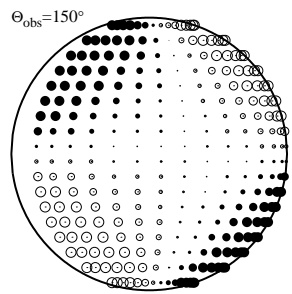
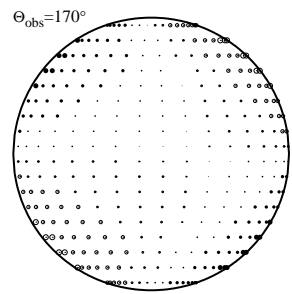
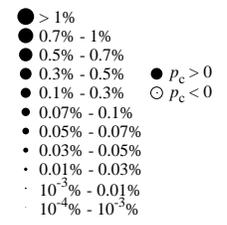
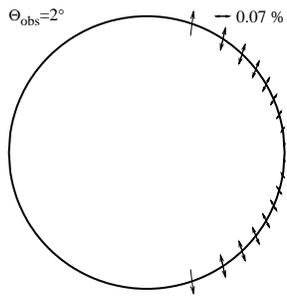
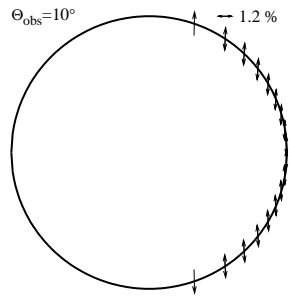
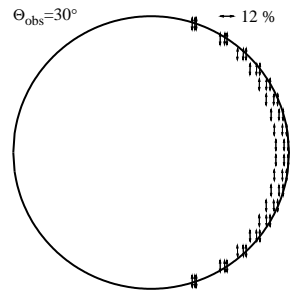
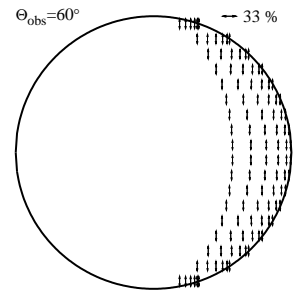
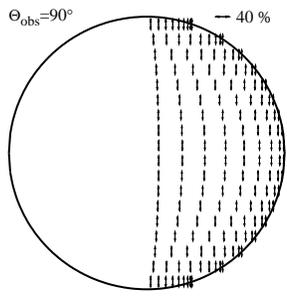
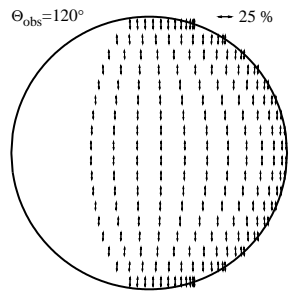
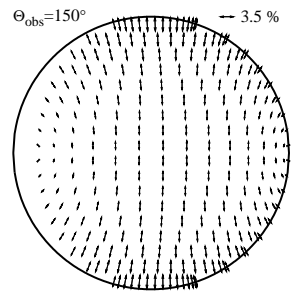
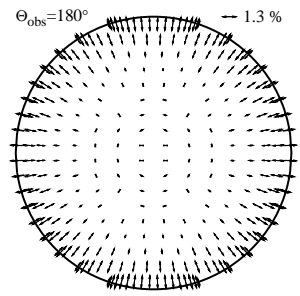
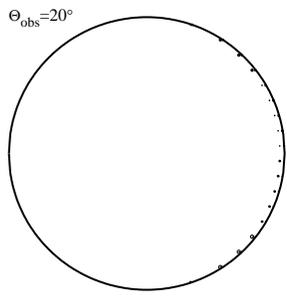
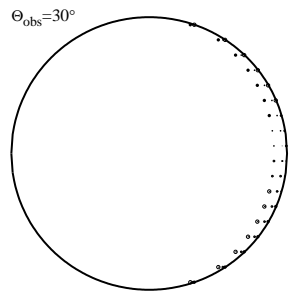
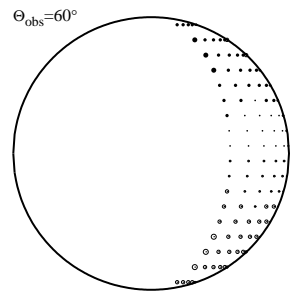
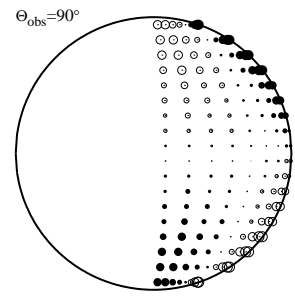
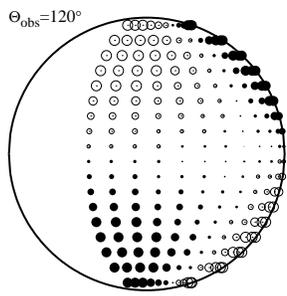
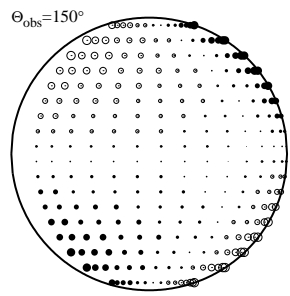
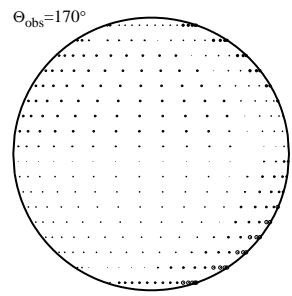
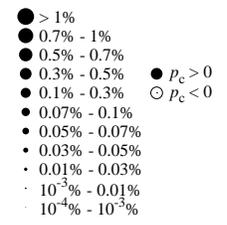
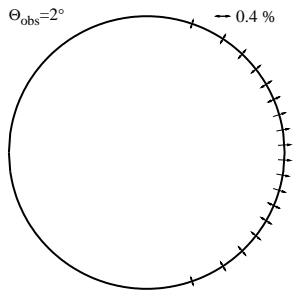
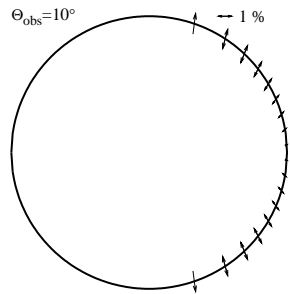
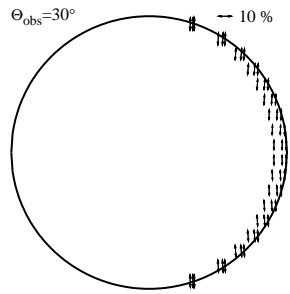
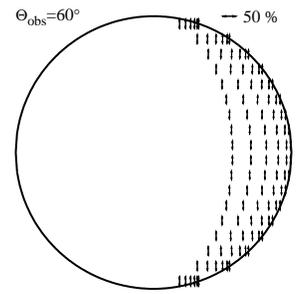
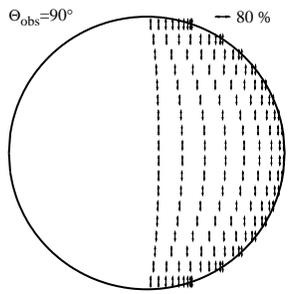
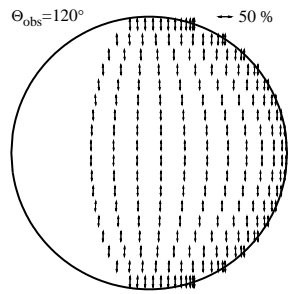
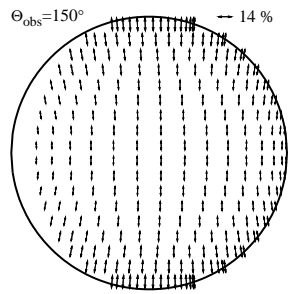
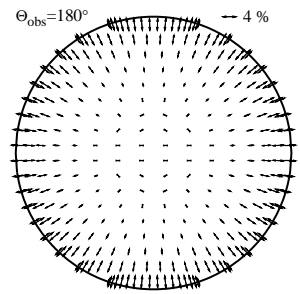
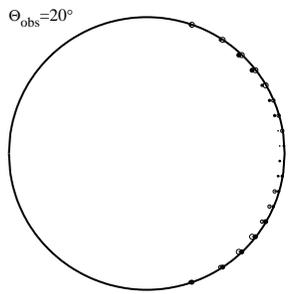
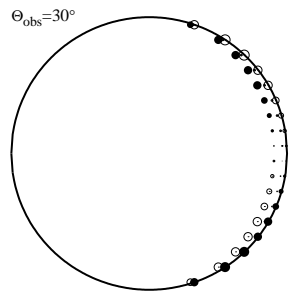
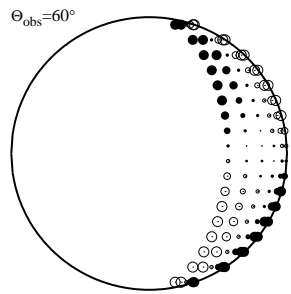
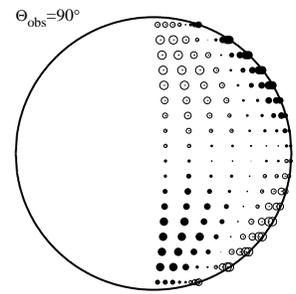
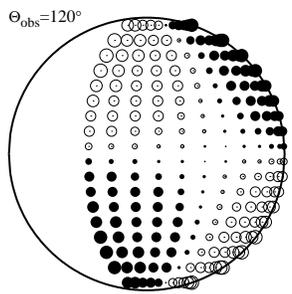
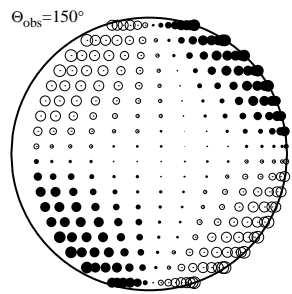
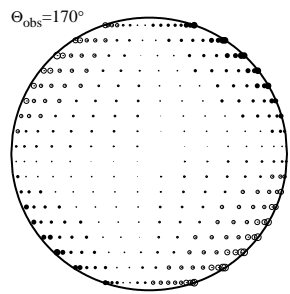
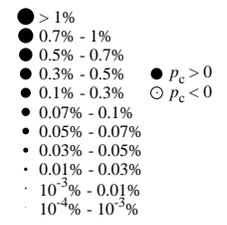
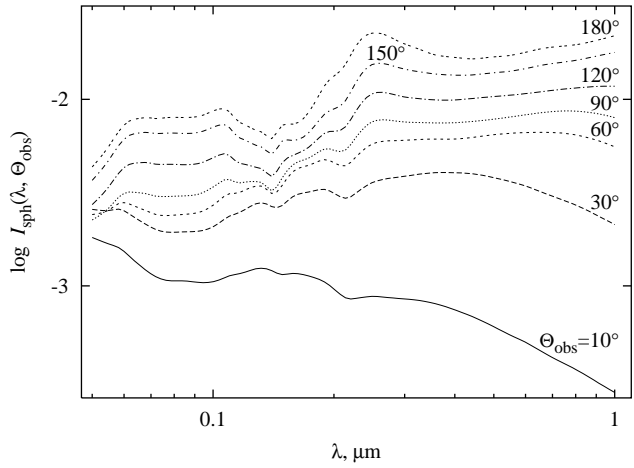
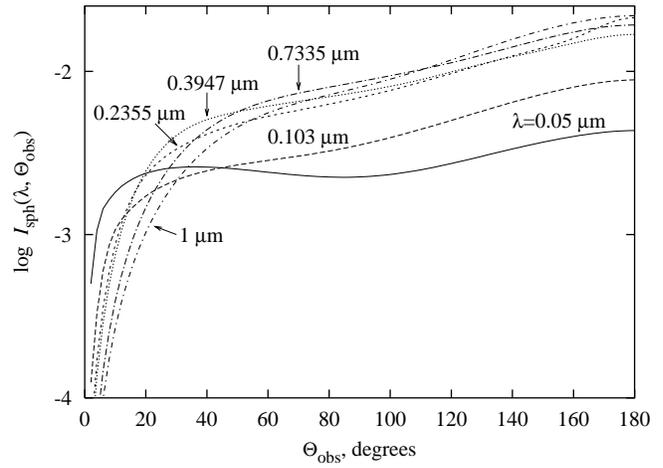
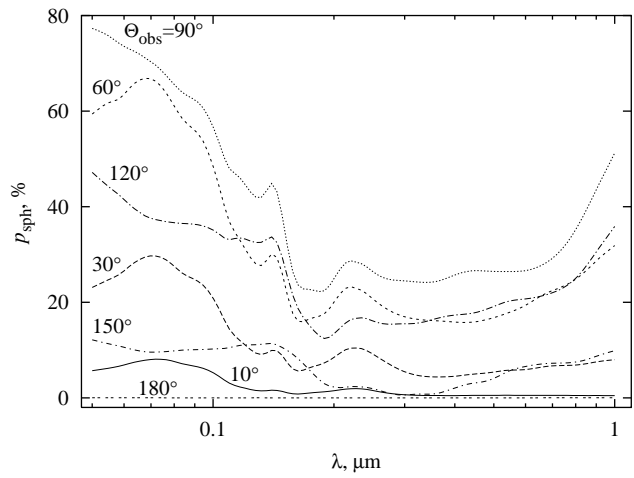
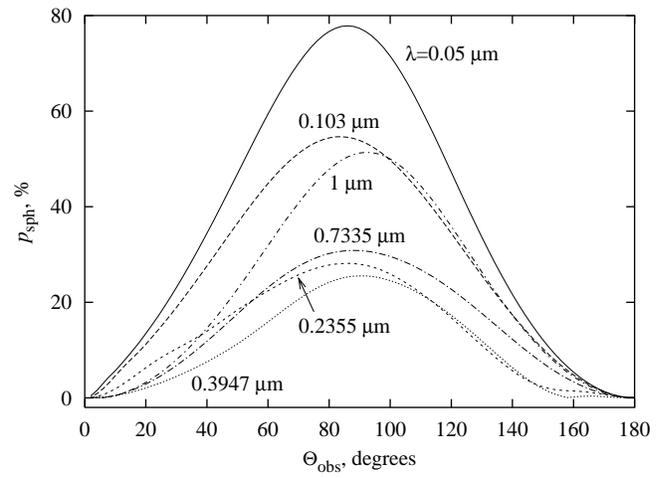
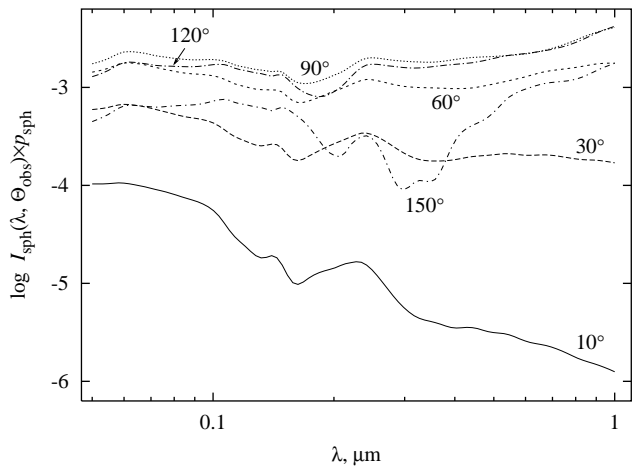
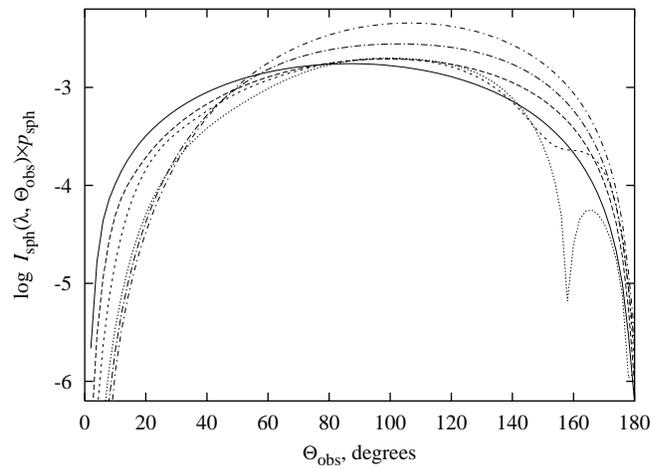
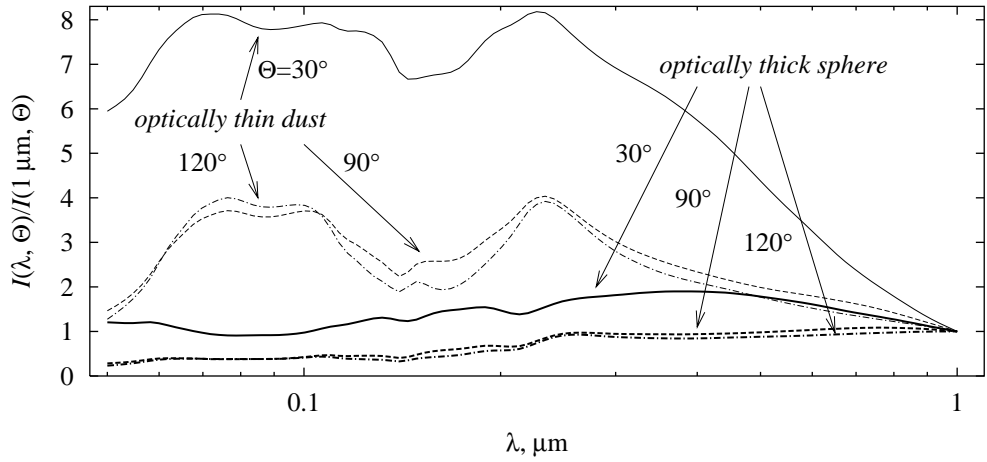
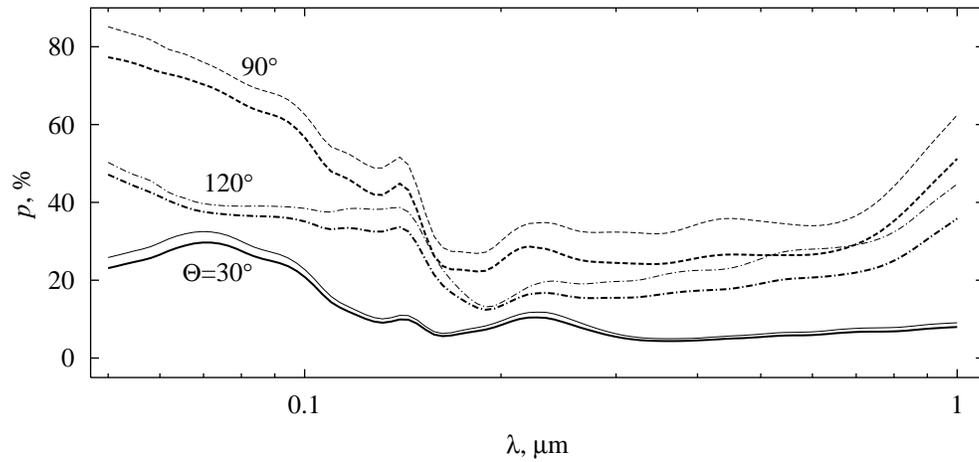
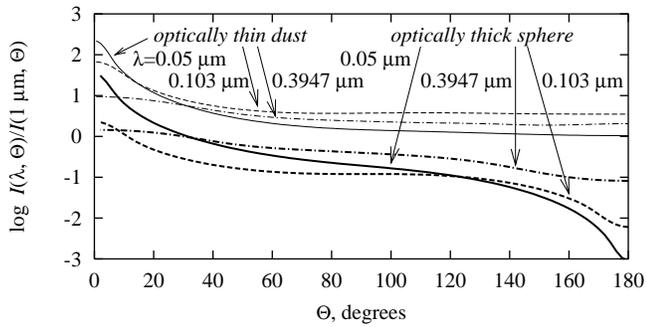
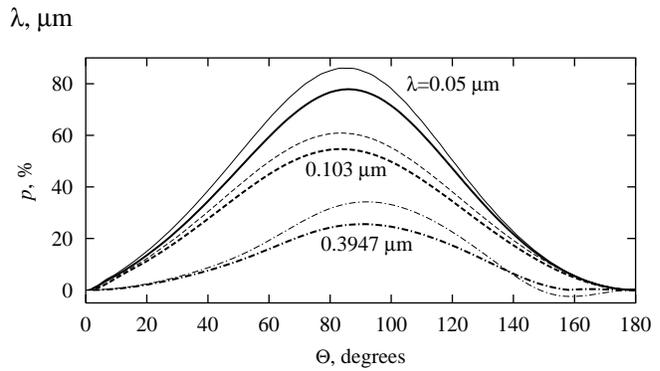
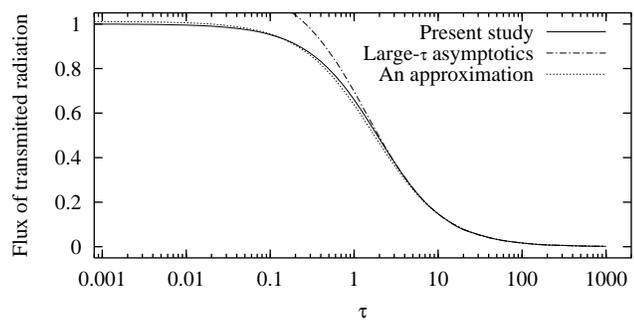
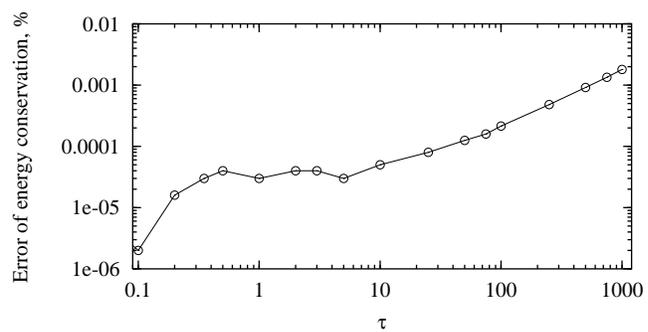
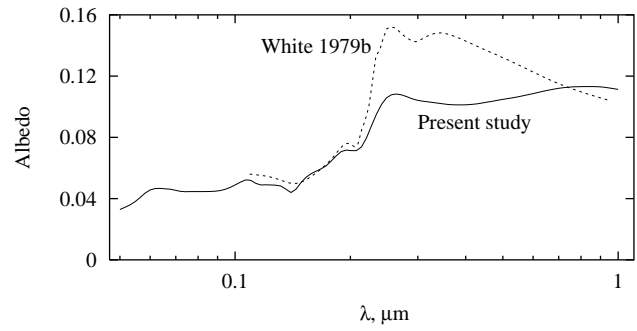
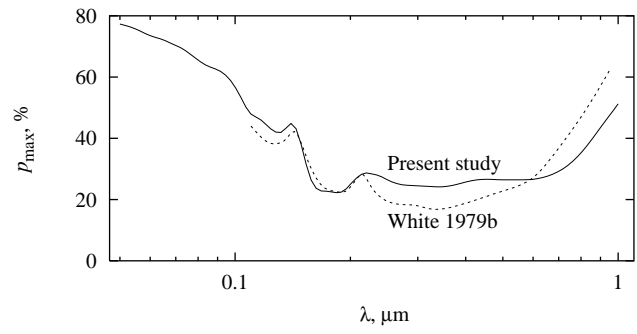
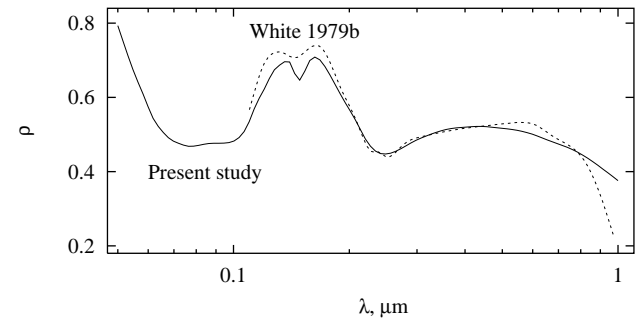
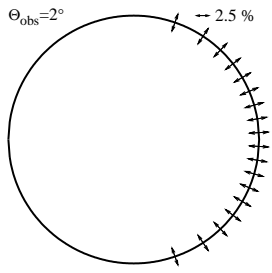
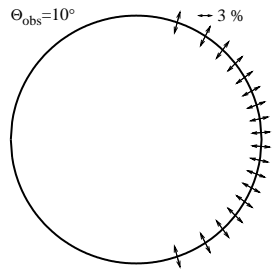
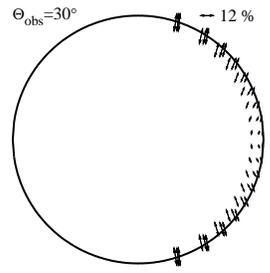
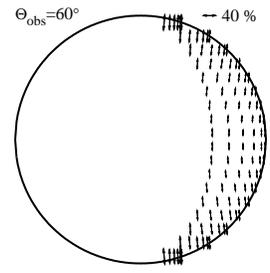
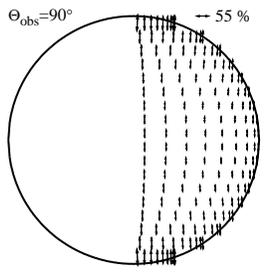
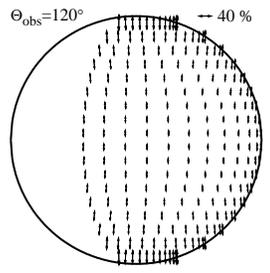
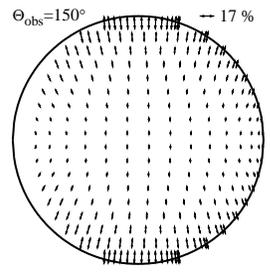
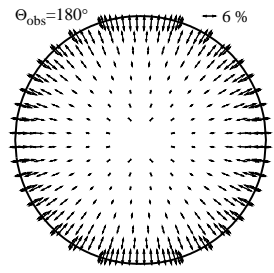
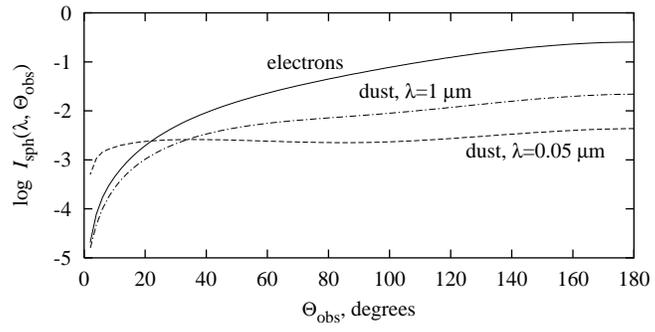
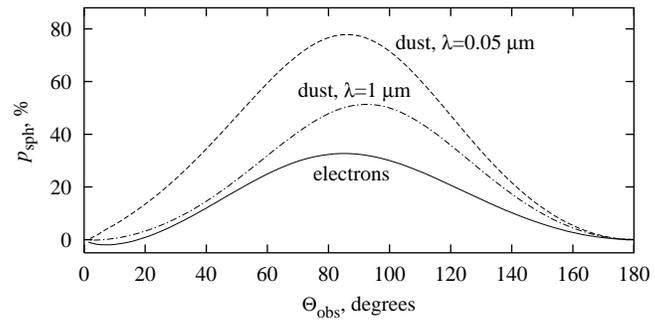
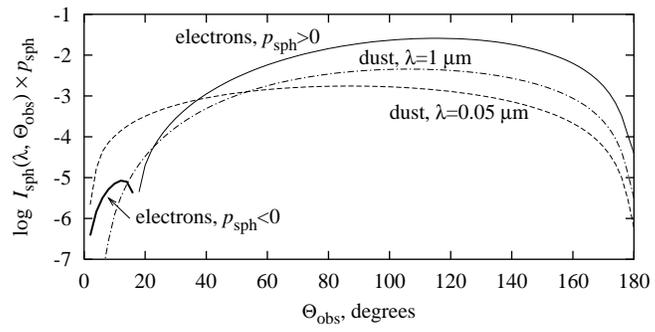
\end{document}